\documentclass[aps,prx,superscriptaddress,twocolumn,longbibliography, notitlepage,nofootinbib]{revtex4-1}

\usepackage{color}
\definecolor{color1}{rgb}{0,0,0.7}
\definecolor{color2}{rgb}{0.85,0,0}

\usepackage[breaklinks=true,colorlinks,citecolor=red,linkcolor=color2,urlcolor=color1]{hyperref}
\usepackage{graphicx}
\usepackage{amsmath}
\usepackage{amssymb}
\usepackage{amsfonts}
\usepackage{amsthm}
\usepackage{bbm}
\usepackage{hyperref}
\usepackage{soul}
\usepackage{textcomp}
\usepackage{bm}
\usepackage{dsfont}
\usepackage{relsize}
\usepackage{braket}
\usepackage{enumitem}
\usepackage{cleveref}
\usepackage{bigints}
\usepackage{float}
\usepackage{framed}
\usepackage[makeroom]{cancel}

\usepackage{pifont}
\newcommand{\cmark}{\text{\ding{51}}}
\newcommand{\xmark}{\text{\ding{55}}}

\theoremstyle{plain}

\numberwithin{obs}{section}
\newtheorem{lem}{Lemma}

\newcommand{\lemref}[1]{\textcolor{color2}{\hyperref[#1]{Lemma~}\bfseries{\ref{#1}}}}

\newcommand{\eref}[1]{\textcolor{color2}{\hyperref[#1]{eq.$\,$(\ref{#1})}}}
\newcommand{\Eref}[1]{\textcolor{color2}{\hyperref[#1]{Eq.$\,$(\ref{#1})}}}
\newcommand{\fref}[1]{\textcolor{color2}{\hyperref[#1]{Fig.$\,$\bfseries{\ref{#1}}}}}
\newcommand{\tref}[1]{\textcolor{color2}{\hyperref[#1]{Table~\bfseries{\ref{#1}}}}}
\newcommand{\eq}[1]{\textcolor{color2}{(\ref{#1})}}

\newcommand{\aref}[1]{\textcolor{color2}{\hyperref[#1]{App.$\,$\ref{#1}}}}

\newcommand{\comments}[1]{}
\newcommand{\ba}{\begin{align}}
\newcommand{\ea}{\end{align}}
\newcommand{\eps}{\varepsilon}

\newcommand{\Tr}[1]{\text{Tr}\left[#1\right]}

\newcommand{\rom}[1]{\uppercase\expandafter{\romannumeral #1\relax}}

\def\Tr{{\rm Tr}}

\newcommand{\nocontentsline}[3]{}
\newcommand{\MT}[2]{\vspace{10pt}\bgroup\let\addcontentsline=\nocontentsline#1{#2}\egroup\vspace{-10pt}}

\begin{document}

\title{Collective advantages in finite-time thermodynamics}

\author{Alberto Rolandi}
\email{alberto.rolandi@unige.ch}
\affiliation{D\'{e}partement de Physique Appliqu\'{e}e,  Universit\'{e} de Gen\`{e}ve,  1211 Gen\`{e}ve,  Switzerland}

\author{Paolo Abiuso}
\email{paolo.abiuso@oeaw.ac.at}
\affiliation{Institute for Quantum Optics and Quantum Information - IQOQI Vienna,
Austrian Academy of Sciences, Boltzmanngasse 3, A-1090 Vienna, Austria}

\author{Martí Perarnau-Llobet}
\email{marti.perarnaullobet@unige.ch}
\affiliation{D\'{e}partement de Physique Appliqu\'{e}e,  Universit\'{e} de Gen\`{e}ve,  1211 Gen\`{e}ve,  Switzerland}

\begin{abstract}
A central task in finite-time thermodynamics is to minimize the excess or dissipated work $W_{\rm diss}$ when manipulating the state of a system immersed in a thermal bath. We consider this task for an $N$-body system whose constituents are identical and uncorrelated at the beginning and end of the process. In the regime of slow but finite-time processes, we show that $W_{\rm diss}$ can be dramatically reduced by considering collective protocols in which interactions are suitably created along the protocol. This can even lead to a sub-linear growth of $W_{\rm diss}$ with $N$: $W_{\rm diss}\propto N^x$ with $x<1$; to be contrasted to the expected~$W_{\rm diss}\propto N$ satisfied in any non-interacting protocol. We derive the fundamental limits to such collective advantages and show that $x=0$ is in principle possible, however it requires long-range interactions. We explore collective processes with spin models featuring two-body interactions and achieve noticeable gains under realistic levels of control in simple interaction architectures. As an application of these results, we focus on the erasure of information in finite time and prove a faster convergence to Landauer's bound.
 \end{abstract}

\maketitle

\MT\section{Introduction}
Collective effects play a central role in physics, ranging from phase transitions to quantum entanglement. Often, they can be exploited for a useful task, such as ultra-precise measurements~\cite{Giovannetti2006}, leading to the notion of a collective advantage\footnote{The outcome of a task is improved when performed globally on a collection of systems than when realized on each system individually.}.
In the growing fields of stochastic and quantum thermodynamics~\cite{Esposito2009,Jarzynski2011,Seifert2012,Goold2016,Vinjanampathy2016,Mitchison2019,myers2022quantum}, such advantages have received notable attention: relevant examples are found in quantum batteries~\cite{Campaioli2018,Bhattacharjee2021,Hovhannisyan2013,Binder2015,Campaioli2017,JuliaFarre2020,Rossini2020,Gyhm2022},  where entangling operations have been proven to enable faster charging~\cite{Campaioli2018,Bhattacharjee2021,Hovhannisyan2013,Binder2015,Campaioli2017,JuliaFarre2020,Rossini2020,Gyhm2022}; 
 in many-body thermal engines~\cite{Mukherjee2021}, whose performance can be enhanced via phase transitions~\cite{campisi2016power,Vroylandt2017,Holubec2017,Herpich2018,Abiuso2020,Fogarty2020,Barra2022}, many-body interactions~\cite{Souza2022Collective,Solfanelli2023}, or superradiance~\cite{Niedenzu_2018,Kloc2019Collective,Kolisnyk2023,Jaseem2023}; and in quantum transport~\cite{Landi2022Nonequilibrium,Brandner2018,Chiaracane2020,Ehrlich2021,Gerry2022Absence,kamimura2022universal}.

In this work, we uncover a new collective advantage in a crucial task in non-equilibrium thermodynamics: the minimization of dissipation in finite time~\cite{Andresen2022,Salamon1983ThermodynamicLength,Schmiedl2007Schmiedl,Esposito2010,Aurell2011,Sivak2012a,Bonana2018,dechant2019thermodynamic,Abiuso2020Geometric,Deffner2020,VanVu2021,VanVu2023Thermodynamic}. In general, the thermodynamic work $W$ required to transform a system, in contact with an environment, in a finite time $\tau$ can be split into two contributions (see e.g.~\cite{Jarzynski2011})
\begin{equation}
    W= \Delta F + W_{\rm diss}
     \label{eq:Wdiss}
\end{equation}
a reversible contribution $\Delta F$, the free energy change, and an irreversible positive contribution $W_{\rm diss}$, the excess or dissipated work (the latter is directly proportional to the entropy production~\cite{Landi2021rev}). Whereas $\Delta F$ is extensive with the size $N$ of the system, we will show here that $W_{\rm diss}$ can grow sub-linearly in $N$. 
This is proven in the regime of slow-but-finite-time processes and becomes possible by exploiting many-body interactions suitably created along the process. 

The advantage is dramatic: in principle, collective processes enable an $N$-fold reduction of $W_{\rm diss}$ when compared to local processes (see \fref{fig}). While we will show that reaching this limit requires highly non-local or long-range interactions, a sub-linear growth of $W_{\rm diss} $ can be achieved with two-body interactions and realistic control. 
 
To obtain these results, we rely on the framework of thermodynamic geometry~\cite{Salamon1983ThermodynamicLength,Crooks, Sivak2012a,scandiThermodynamicLengthOpen2019}, which has recently found numerous applications in mesoscopic and quantum  systems~\cite{Zulkowski2012,Rotskoff2017,Abiuso2020,Brandner2020,Li2022,Eglinton2022,Frim2022,Alonso2022Geometric,Mehboudi2022Thermodynamic,Eglinton_2023}. 
In this approach, which is valid in the slow-driving regime, finite-time protocols are identified with curves in the thermodynamic parameter space, so that geodesics are those protocols that minimize $W_{\rm diss}$. Our results show that geodesic protocols generically explore highly interacting Hamiltonians, even if interactions are absent at the beginning and end of the process. As an application, we focus on finite-time information erasure~\cite{Zulkowski2014,Proesmans2020,boyd2022shortcuts,Lee2022,Diana2013,Zhen2021,Zhen2022,Ma2022,Scandi2022,Konopik2023,rolandi2022finite} of $N$ qubits. We show that collective processing can substantially reduce dissipation in this relevant task, leading to a faster convergence to Landauer's bound.

Overall, these results uncover a genuine collective advantage in stochastic and quantum thermodynamics, which is not linked to standard collective phenomena such as quantum entanglement, phase transitions, or collective system-baths couplings (e.g. superradiance).

\MT\section{Framework}
Let us consider a system in a $d$-dimensional Hilbert space $\mathbb C^d$ with an externally driven Hamiltonian~$\hat h(t)$. It can be parameterized as $\hat h(t)=\sum_{j=1}^n \lambda^j(t) \hat x_j$, $\{ \lambda^j \}$ are externally controllable parameters, and $\{\hat x_j \}$ are the corresponding observables. These control parameters can be constrained, we will denote by~$M\subseteq \mathbb{R}^n$ the manifold of the allowed values for these parameters. Additionally, the system is in contact with a thermal bath at inverse temperature~$\beta$. 

We focus on the task of driving $\hat h(t)$ from an initial configuration $\hat h(0) =\hat h_A$ to a final one $\hat h(\tau)= \hat h_B$ in a time~$t\in [0,\tau]$. External energy is needed to realize this transformation, quantified by the (average) thermodynamic work: 
\begin{equation}
    W= \int_0^\tau\!\!dt~ \Tr\Big[\frac{d\hat h(t)}{dt} \hat\rho(t)\Big]~,
\end{equation}
where $\hat\rho(t)$ is the state of the system. This expression can be split as in \eref{eq:Wdiss}, where $\Delta F= \beta^{-1} \ln \mathcal{Z}(0)/ \mathcal{Z}(\tau)$ and~$\mathcal{Z}(t)= \Tr\big[e^{-\beta \hat h(t)}\big]$. Whereas 
$\Delta F$ depends only on the endpoints of the process, $W$ depends on the protocol, i.e. the specific driving $\lambda : [0,\tau]\rightarrow M$. The minimal dissipated work $W_{\rm diss}$ in a finite time $\tau$ can then be found by optimising for $\{ \lambda_j(t) \}$ over the space of curves in $M$ connecting $\hat h_A$ to $\hat h_B$: $W_{\rm diss}^{*} \equiv \min_{\lambda\in \mathcal C_{A,B}(M)} W_{\rm diss}$.
To address the non-trivial optimization we make some assumptions. 

First, we assume that the driving $\frac{d}{dt}\hat h(t)$ is slow compared to the relaxation rate. 
Then $W_{\rm diss}$ can be expressed as a quadratic form at leading order in $\tau^{-1}$~\cite{Salamon1983ThermodynamicLength,Crooks, Sivak2012a,Abiuso2020}: 
\begin{align}
   W_{\rm diss}= k_B T \int_0^\tau dt~\dot{\lambda}^i(t)\dot{\lambda}^j(t) g_{ij}(\lambda(t)) +  \mathcal{O}\!\left(\tau^{-2}\right) ~,
    \label{eq:GeneralExpansionDiss}
\end{align}
where $g_{ij}(\lambda)$ is the so-called thermodynamic metric, and we adopted the Einstein summation convention. The metric allows us to define the length of a line element in $M$ by $ds^2 = g_{ij}d\lambda^id\lambda^j$, which is used to assign a length to a curve $\lambda$ in $M$: $L[\lambda] = \int_\lambda \!ds = \int_0^\tau \!dt \sqrt{ \dot{\lambda}^i(t)\dot{\lambda}^j(t) g_{ij}(\lambda(t))}$. It is related to the dissipated work via a Cauchy-Schwartz inequality~\cite{Salamon1983ThermodynamicLength}: $\beta W_{\rm diss}  \geq L^2/\tau$, where equality is satisfied by protocols with constant dissipation rate. The shortest length $\mathcal L$ corresponds to the protocol that minimizes dissipation: $\beta W_{\rm diss}^{*}  = \mathcal L^2/\tau$. We can then find $W_{\rm diss}^{ *}$ by solving the geodesic equation for the thermodynamic metric~\cite{Salamon1983ThermodynamicLength, Sivak2012a,Abiuso2020}.

As a second simplification, we assume there is a single relaxation timescale $\tau_{\rm eq}$\footnote{We assume that all driven observables decay exponentially to equilibrium with the same timescale \cite{Abiuso2020,Abiuso2020Geometric}.}, so that the metric becomes~\cite{scandiThermodynamicLengthOpen2019}: 
\begin{equation}
\label{eq:metric}
    g_{ij}  = \tau_{\rm eq}\frac{\partial^2 \ln \mathcal{Z} }{\partial \lambda^i \partial \lambda^j }~.  
\end{equation}
Note that $g_{ij}$ then becomes the standard thermodynamic metric for macroscopic systems~\cite{Schlogl1985,Salamon1985,Crooks}, which can also describes step-processes~\cite{Nulton1985,Scandi2020}. In what follows, without loss of generality, we set $\tau_{\rm eq}=1$.

As a last simplification, we will assume that the initial and final Hamiltonian commute $[\hat h_A, \hat h_B] = 0$. This allows us to conclude that at all times $[\frac{d}{dt}\hat h(t), \hat h(t)] = 0$, as changes in the eigenbasis can only increase dissipation in the linear response regime~\cite{Abiuso2020Geometric,Brandner2017}.

Let us now consider a scenario in which we perform the driving on $N$ copies of the system. We denote by $\hat H(t) = \hat H_0(t) + \hat H_{\rm int}(t)$ the total Hamiltonian for all the copies, where $\hat H_0(t) = \sum_{j=1}^N \hat h^{(j)}(t)$ and $\hat H_{\rm int}(t)$ contains the interaction between the copies. We parameterize $\hat H(t)$ similarly to $\hat h(t)$: $\hat H(t) = \sum_{i=1}^n \gamma^i(t)\hat X_i$, where the sum can have up to $n=d^N$ terms. The problem at hand imposes the following boundary conditions on the protocol: $\hat H_{\rm int} (0) = \hat H_{\rm int} (\tau) = 0$, $\hat h^{(j)}(0) = \hat h_A$, and $\hat h^{(j)}(\tau) = \hat h_B$ $\forall j$. Furthermore, by the same reasoning as in the case for a single copy, we have that $[\frac{d}{dt}\hat H(t), \hat H(t)] = 0$ for the geodesic protocol.

\MT\section{Fundamental limit of collective advantages}
Let us first note that $\Delta F$ is extensive with $N$ which directly follows from the boundary conditions. Instead, $W_{\rm diss}$ depends on the process and can exhibit a non-trivial behavior whenever $ \hat H_{\rm int} (t) \neq 0$. Indeed, we find that, in general, geodesic paths explore highly interacting Hamiltonians if the constraints allow for it.

To reach the fundamental limit of $W_{\rm diss}^{*}$ we can assume full control on $\hat H(t)$, so that the $n = d^N$ different eigenenergies $\{\gamma^i\}$ can be externally controlled at will --the corresponding $\{\hat X_i\}$ are chosen to be the corresponding projectors. In this case, the thermodynamic metric is given by $\beta^{-2}g_{ij} = \omega_i \delta_{ij} - \omega_i\omega_j$,
where $\omega_i = e^{-\beta\gamma_i}/\mathcal Z$ are the eigenvalues of the thermal state $\hat\omega_\beta = e^{-\beta \hat H}/\mathcal Z$. For which the distance function is known to be the quantum Hellinger angle: $\mathcal{L} = 2 \arccos\Tr\big[\sqrt{\hat\omega_{\beta}(0)} \sqrt{\hat\omega_{\beta}(\tau)}\,\big]$ (cf.~\aref{A} and~\cite{Jencova2004,Abiuso2020}), leading to: 
\begin{equation}
     \beta W_{\rm diss}^{*}= \frac{1}{\tau}\left( 2 \arccos\Tr\!\left[\sqrt{\hat\omega_{\beta}(0)} \sqrt{\hat\omega_{\beta}(\tau)}\right]\right)^2 ~.
     \label{fundamental_bound}
\end{equation}
Since trivially $\arccos (x) \leq \pi/2$ for $x>0$, the minimal dissipation of a $N$-body system is bounded by a constant $W_{\rm diss}^{*} \leq \frac{1}{\tau} \pi^2$ independent of $N$. This is somehow astonishing, as we expect the dissipation generated when driving a many-body system to increase extensively with its size. In~\aref{A} we derive the protocol that achieves this limit, obtaining
\begin{equation}\label{FC_driving}
    \beta\hat H(t) \!=\! -2\log\!\left[\sin\!\!\left[\frac{\mathcal L(\tau\!-\!t)}{2\tau}\right]\!\!\sqrt{\hat \omega_\beta(0)} + \sin\!\!\left[\frac{\mathcal Lt}{2\tau}\right]\!\!\sqrt{\hat \omega_\beta(\tau)} \right]\!.
\end{equation}
Crucially, this protocol generally requires all possible interacting terms available in the Hamiltonian space, including highly non-local $N$-body interactions (see proof in \aref{A}). This is illustrated in what follows for the paradigmatic task of erasing $N$ bits of information.

\MT\section{Collective erasure}
Let us consider $N$ qubits, each with local Hamiltonian $\hat h(t)=\eps(t) \hat\sigma_z$. We want to drive $\eps(t)$ from $\eps(0) = 0$ to $\eps(\tau) = E$ with $E\gg k_B T$, so that the state of each qubit evolves from a fully mixed state $\hat\omega_\beta(0)=\frac{1}{2}\mathbbm{1}$ to an (almost) pure state $\hat\omega_\beta(\tau) \approx \ket{0}\!\!\bra{0}$ due to the action of the external bath. We have $\Delta F = N k_B T \ln 2$, corresponding to Landauer's bound.

Consider first the independent scenario, so that during the whole protocol $\hat H_{\rm int}(t) = 0$. For each qubit, the dissipation generated via an optimal driving can be computed from~\eref{fundamental_bound} with the aforementioned boundary conditions, yielding $\beta W^{*}_{\rm diss} = \pi^2/4\tau $, see also~\cite{Salamon1983ThermodynamicLength,Ma2022,rolandi2022finite}. The total dissipation of $N$ qubits then reads:
\begin{align}
     \beta W_{\rm diss}^{\rm *,  local} = N \frac{\pi^2}{4 \tau}~, 
     \label{localbound}
\end{align}
which  grows linearly with $N$.
The corresponding optimal driving reads $\beta\eps(t)=\ln\tan\!\left[\pi(t+\tau)/4\tau\right]$, which has been implemented experimentally in a single-energy driven dot~\cite{Scandi2022}. 

\begin{figure*}[ht]
    \centering
    \includegraphics[width=0.99\textwidth]{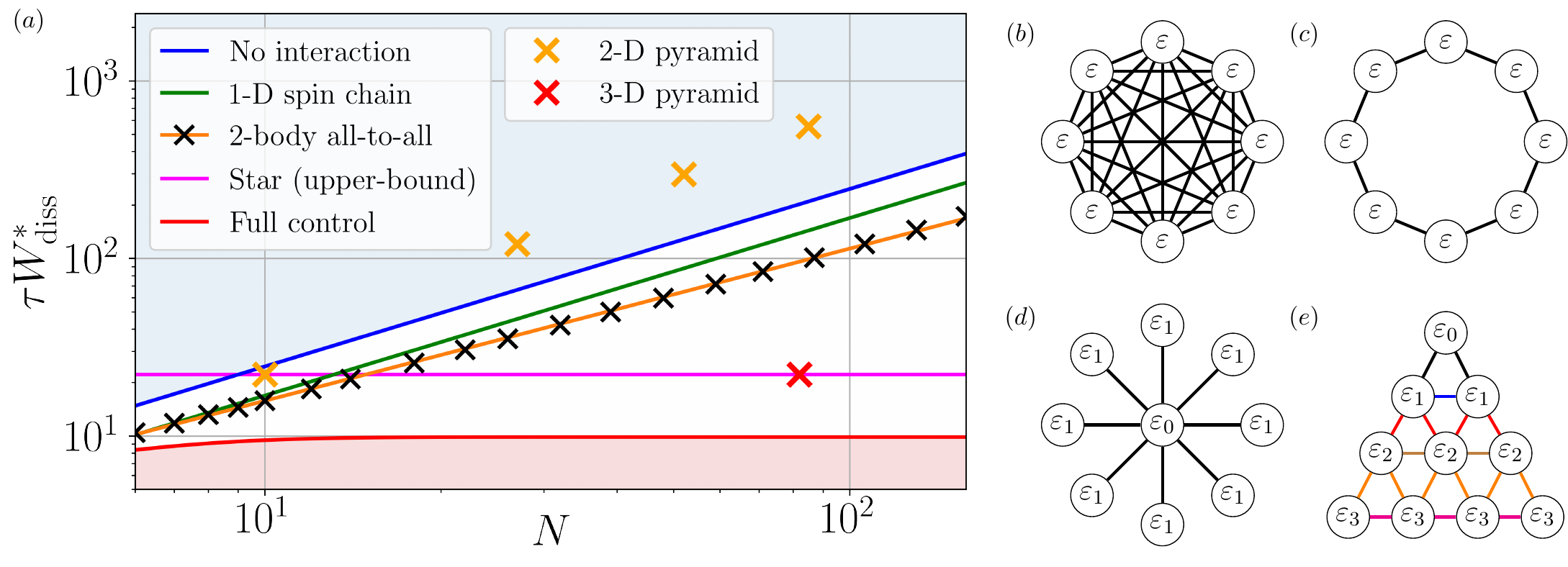}
    \vspace{-15pt}
    \caption{(a) Minimal dissipation for the erasure of $N$ spins for different control designs analyzed in this work. These are compared with the dissipations that are achievable with no interactions (\eref{localbound}, blue-shaded area), and with the dissipations that are not achievable regardless of the protocol (\eref{boundNqubits}, red-shaded area). We find $\tau W_{\rm diss}^{\rm *, chain}\approx 1.69 N$,  $\tau W_{\rm diss}^{\rm *, all} \approx 2.20 N^{0.857}$, while $\tau W^{\rm *, Star}_{\rm diss} \leq 9\pi^2/4$ (cf.~\aref{B}). Single points are provided for 2-D and 3-D Pyramids with few layers and an aperture of 8 (cf.~\aref{B}). (b-e) Depiction of the geometries of the interactions in \eref{eq:2B_hamiltonian} (equal colors/labels correspond to equal values of the local fields). (b) all-to-all model with $N=8$, (c) 1-D spin chain with $N=8$, (d) the Star model with $N=9$, (e) 2-D Pyramid model with $4$ layers and an aperture of $1$.}
    \vspace{-17pt}
    \label{fig}
\end{figure*}

If we now allow for full control of the Hamiltonian, we can again use~\eref{fundamental_bound} to compute the minimal dissipation, but this time we use the global states $\hat\omega_\beta(0)= \frac{1}{2^N} \mathbbm{1}$ and $\hat\omega_\beta(\tau) \approx \ket{0}\!\!\bra{0}^{\otimes N}$ instead of the local ones. This leads to: 
\begin{equation}
    \beta W_{\rm diss}^{\rm *, global}  = 
     \frac{1}{\tau}\left(2\arccos\!\left[\frac{1}{2^{N/2}}\right] \right)^2 
     = \frac{\pi^2}{\tau} + \mathcal{O}\!\left( 
     e^{-N/2} \right)~. 
     \label{boundNqubits}
\end{equation}
Therefore, an $N$-fold advantage can potentially be achieved by global processes, as illustrated in \fref{fig}. 

Let us now discuss the implications of this result for the reachability of Landauer's bound. From \eref{eq:Wdiss} we have $\Delta F = N k_B T \ln 2$ whereas $W_{\rm diss}$ can reach \eref{boundNqubits} at leading order in $\tau^{-1}$ (recall that our results are based on the slow driving assumption where the expansion~\eref{eq:GeneralExpansionDiss} is well justified). Hence, the work cost per qubit of erasure can be written as:
\begin{equation}
    \beta W_{\rm qubit}^* = \ln 2 + \frac{\pi^2}{\tau N} +\mathcal{O}\!\left(\tau^{-2}\right).
    \label{eq:ReachingLandauer}
\end{equation} 
Hence, in the thermodynamic limit $N\rightarrow \infty$, we can approach Landauer's bound with an error that scales as $\tau^{-2}$ instead of the standard $\tau^{-1}$~\cite{Zulkowski2014,Proesmans2020,Ma2020Experimental,boyd2022shortcuts,Lee2022,Diana2013,Zhen2021,Zhen2022,Ma2022,Scandi2022,Konopik2023,rolandi2022finite}.

We note that a collective $N$-qubit erasure based upon spontaneous symmetry breaking has been recently reported in Ref.~\cite{Buffoni2023}. Furthermore, a link between complexity, as in higher level $k$-body interactions and faster information erasure has been suggested in Ref.~\cite{Taranto2023}. While being based on different collective phenomena or operations, these results share conceptual similarities to the collective erasure developed here.

The optimal driving achieving the limit \eref{boundNqubits} can be computed from \eref{FC_driving}: 
\begin{equation}
\label{eq:Nbody_prot}
    \beta \hat H (t)= \gamma(t)\sum_{j=1}^N (-1)^{j+1}\!\! \hspace{-4mm} \sum_{i_1<i_2<...<i_j}^N \hspace{-4mm} \hat x^{(i_1)} \hat x^{(i_2)} ... \hat x^{(i_j)}~,
\end{equation}
where $\hat x = \hat\sigma_+\hat\sigma_-$ and the control function can be written as $\gamma(t) = 2\log\!\left[1+ 2^{N/2}\sin\!\left(\frac{\pi t}{2\tau}\right)\sin\!^{-1}\!\left(\frac{\pi(\tau-t)}{2\tau}\right)\right]$. It follows that highly non-local $N$-body interactions are required to saturate the bound~\eref{boundNqubits}. More specifically, one needs to activate every possible (classical) interaction present in the system. This makes reaching the fundamental bound \eref{boundNqubits} highly challenging in practice, and opens the question as to whether collective advantages beyond the local bound \eref{localbound} can be achieved via more realistic driven many-body systems featuring (local) few-body interactions. We address this relevant question in what follows. 

\MT\section{Collective advantage in driven many-body systems}
To seek collective advantages in a more realistic model, in this section, we constrain the total system to only feature at most 2-body interactions. Specifically, we consider a spin system with Hamiltonian of the form
\begin{equation}\label{eq:2B_hamiltonian}
\hat H(t) = \sum_{i=1}^N\eps_i (t)\hat\sigma_z^{(i)} + \frac{1}{2}\sum_{i,j=1}^N J_{ij}(t) \hat\sigma_z^{(i)}\hat\sigma_z^{(j)}    
\end{equation}
We thus examine different degrees of control, reflected in the topologies represented in \fref{fig}: (i) an all-to-all spin model, (ii) a 1-D spin chain with nearest neighbor interaction (with periodic boundary conditions) (iii) a Star-shaped design, which we generalize to (iv) a multi-layer Pyramid scheme.
In practice, the energies $\eps_i(t)$ could be tuned via an external magnetic field whereas the interaction strength $J_{ij}(t)$ could be controlled by changing the distance between the spins interacting via dipole-dipole coupling. Current quantum annealers have the capacity of tuning generic Hamiltonians of the form~\eref{eq:2B_hamiltonian}~\footnote{See for example \href{https://www.dwavesys.com/}{D-Wave Systems}.}.

The all-to-all model corresponds to taking uniform magnetic fields and  spin interactions, i.e. $\eps_i(t)\equiv\eps(t)$ and $J_{ij}(t)\equiv J(t)$ in \eref{eq:2B_hamiltonian}. We can compute the partition function as follows 
\begin{equation}
    \mathcal Z_{\rm all} = \sum_{k=0}^N \binom{N}{k} e^{-\beta E_k}~,
\end{equation}
where $E_k = \eps(2k-N) + \frac{1}{2}J(2k-N)^2$. The standard 1-D Ising model corresponds to uniform local terms $\eps$, and $J_{i,i+1}\equiv J$ for nearest neighbours and $0$ elsewhere. The partition function can be found by making use of the transfer matrix method:
\begin{align}
\mathcal{Z}_{\rm chain} &= z_+^N + z_-^N\;,
\end{align}
where $z_\pm =e^{-\beta J/2} \cosh{\beta \eps}\pm\sqrt{e^{-\beta J} \sinh{\beta \eps} +e^{\beta J} }$.
Thirdly, we consider a Star topology corresponding to a central spin $\hat\sigma_z^{(1)}$ with local magnetic field $\eps_0(t)$ and uniform elsewhere $\eps_i(t)\equiv\eps_1(t)\; \forall\, i>1$, and uniform ``radial'' interaction $J_{1j}(t)\equiv J(t)$, and $0$ elsewhere. The partition function is easily computed as
\begin{align}
    \mathcal{Z}_{\rm Star} & =e^{-\beta\eps}(2\cosh{\beta\lambda_+})^{N-1}+e^{\beta\eps}(2\cosh{\beta\lambda_-})^{N-1}\;,
    \label{eq:ZStar}
\end{align}
where $\lambda_{\pm} =\eps_1\pm J$.

For the models above,  given the partition function, we compute the metric according to \eref{eq:metric}, from which we can obtain the geodesic equations. Their solution provides us with the minimal length for given boundary conditions, from which we find the minimal dissipation. This is implemented numerically for the task of approximate erasure (see details in  \aref{C}), we take $\eps(0) = 0$ and $\eps(\tau) = 5k_BT$ (recall $J(0) = J(\tau) = 0$) which corresponds to an erasure process with an error of $4.5\cdot 10^{-5}$.

In \fref{fig} we present the resulting minimal dissipation for the different many-body models. The results are contrasted with the optimal non-interacting protocol \eref{localbound} and the fundamental bound obtained with full-control \eref{boundNqubits} (i.e. arbitrarily complex interactions). 

First, we observe that the nearest neighbor model displays a linear increase of the dissipation with $N$, but with a better pre-factor than the non-interacting case ($W_{\rm diss}^{\rm *, chain}/W_{\rm diss}^{\rm *, local} \approx 0.686$). On the other hand, the all-to-all model displays a sub-linear dependence on $N$: $W_{\rm diss}^{\rm *, all} = \alpha N^x$ with $x\approx 6/7$. Furthermore, the exponent $x$ displays a non-trivial dependence on the specific boundary conditions (cf. \aref{C}).
Finally, quite remarkably,  the Star model can achieve \emph{a finite value of the dissipation, independent of~$N$}. This feature is enabled by a 3-step protocol (cf. \aref{B}) that suppresses specific terms in the otherwise-extensive $\log{\mathcal{Z}_{\rm Star}}$. Interestingly, the Star model was found to be optimal in the context of two-body probes used for thermometry~\cite{abiuso2022discovery}.

The sub-linearity of the all-to-all's and Star-model's dissipation is remarkable as it allows for the same effect as in \eref{eq:ReachingLandauer}: it is possible to reach Landauer's bound in finite-time with an error that scales as $\tau^{-2}$ instead of $\tau^{-1}$ as one approaches the thermodynamic limit. However, both these models use long-range interactions between arbitrarily far spins as $N$ grows, and their scaling properties might thus be seen as inconsequential. For this reason, in~\aref{B} we generalized the Star model to a multi-layer structure, i.e. a Pyramid model (see \fref{fig} and \aref{B}). By generalizing the Star protocol, we prove that such model can achieve $W_{\rm diss}^{\rm Pyr}\propto \ell^2$, $\ell$ being the number of layers of the pyramid. Given that $N\propto \ell^D$ for pyramids in $D$ dimensions, $W_{\rm diss}^{\rm Pyr}\propto N^{2/D}$ follows asymptotically.

\begin{table}
\centering
\begin{tabular}{c|c|c|c|c}
Model & 1D chain & All-to-All & \ Star \ & Pyramid\\
\hline \hline
Asymptotic $W_{\rm diss}$ & $\mathcal{O}(N)$  & $\mathcal{O}(N^x)$  & $\mathcal{O}(1)$ & $\mathcal{O}(N^{2/D})$ \\  \hline 
  Short-range   & $\cmark$ & $\xmark$  & $\xmark$ & $\cmark$ \\ \hline 
\end{tabular}
\vspace{-5pt}
\caption{All models studied in this work are based on two-body interactions \eq{eq:2B_hamiltonian}. The All-to-all and Star model feature long-range interactions that enable a sub-linear scaling of $W_{\rm diss}$ , i.e. a \emph{collective advantage}. The Pyramid models can achieve such advantage in $D=3$ spatial dimensions using short-range interactions only (cf. \aref{B}).}
\vspace{-25pt}
\label{tab:models}
\end{table}

\MT\section{Conclusions and discussion}
In this work, we considered the task of minimizing dissipated work, $W_{\rm diss}$, for an $N$-body system. We showed that, in contrast to~$\Delta F$, $W_{\rm diss}$ can grow sublineraly with $N$ by suitably creating interactions between the $N$ systems along the process. This leads to a finite-time reduction of dissipation induced by collective processes and has a clear potential for improving various thermodynamic tasks ranging from quantum/stochastic engines~\cite{Seifert2012,Blaber2020,myers2022quantum} to the estimation of equilibrium free energy via non-equilibrium work measurements~\cite{Blaber2020Skewed}; or, as is shown here, for the erasure of information in finite time. There are several observations to be made about these collective processes. 

First, the derived collective processes are a genuine effect of finite-time thermodynamic protocols, which cannot be directly linked to other well-known collective phenomena such as entanglement, phase transitions, or superradiance. Indeed, (i) they do not require the presence of quantum correlations or coherence, but rather arise due to the interplay between interactions and dissipation to an external thermal environment; and (ii) they are process dependent -- i.e. depend on the whole driving protocol~$\hat H(t)$ -- unlike phase transitions which take place in a particular point in the parameter space.

Second, our results suggest an interesting interplay between the complexity of the interactions and the associated reductions in dissipation, see also Ref.~\cite{Taranto2023}. In particular, we argued that reaching the maximal advantage requires highly non-local $N$-body interactions. 
Despite this, we showed that similar reductions (in scaling)  can  be achieved with only two-body long-range interactions via the Star model. A sub-linear growth of $W_{\rm diss}$ was found in the all-to-all model and, crucially, in the Pyramid model that only features short-range strong interactions. See \tref{tab:models} for a compact summary.

Third, being derived in the linear response regime, the dissipated work is directly related to the work fluctuations $\sigma_W^2$ via the work fluctuation-dissipation relation $\frac{\beta}{2} \sigma_W^2 = W_{\rm diss}$ \cite{hermans_simple_1991,jarzynski_nonequilibrium_1997,Mandal2016a,Miller2019}. This  implies that the collective gains also lead to a reduction of work fluctuations, a desired property  in stochastic thermodynamics. 

Finally, it is important to stress that our results have been derived in the slow driving regime, i.e., for the leading order contribution of $W_{\rm diss}$ in $\tau^{-1}$. For a finite (large) time $\tau$, the next order contributions of $\mathcal{O}(\tau^{-2})$ can become relevant when increasing $N$. An exciting future endeavor is to generalize such collective advantages for arbitrary non-equilibrium protocols. For this, it might be useful to exploit recent results on minimal dissipation and the Wasserstein distance~\cite{dechant2019thermodynamic,Chen2020,Nakazato2021,Dechant2022,VanVu2021,Abiuso2022Thermodynamics,VanVu2023Thermodynamic} as well as new tools such as reinforcement learning~\cite{Erdman2022,Erdman2023Pareto} or fast-driving expansions~\cite{Blaber2021,Erdman2023Pareto,Rolandi2022Fast} for finding optimal  protocols.

Another future challenge is to understand how the collective advantages are modified beyond the simple model of thermalization used in \eref{eq:metric} and by adding constraints on the strength of the couplings in \eref{eq:2B_hamiltonian}. In particular, for more realistic thermalization models where the relaxation timescale(s) is modified in the presence of interactions, which can lead, e.g., to a critical slowdown of relaxation.
\vspace{10pt}
\begin{acknowledgments}
\vspace{-10pt}
We warmly thank Nicolas Brunner, Harry J. D. Miller, and Maja Milas for their insightful discussions. This work was supported by the Swiss National Science Foundation through an Ambizione Grant No. PZ00P2-186067. P.A. is supported by the QuantERA II programme, that has received funding from the European Union’s Horizon 2020 research and innovation programme under Grant Agreement No 101017733, and from the Austrian Science Fund (FWF), project I-6004.
\end{acknowledgments}

\bibliographystyle{apsrev4-1}
\bibliography{mybib.bib}

\widetext
\appendix

\section{Full control scenario}\label{A}
We consider $N$ copies of a $d$-dimensional system, with an externally driven Hamiltonian $\hat H(t) = \hat H_0(t) + \hat H_{\rm int}(t)$, where $\hat H_0(t) = \sum_{j=1}^N \hat h^{(j)}(t)$ (with each $\hat h^{(j)}$ acting only on the $d$-dimensional Hilbert space of the $j$-th site) and $\hat H_{\rm int}(t)$ contains the interaction terms. We can always parameterize the total Hamiltonian by $\hat H(t) = \sum_{i=1}^n \gamma^i(t)\hat X_i$ with $n=d^{2N}$.

We focus on the task of driving each copy from an initial configuration $\hat h(0) =\hat h_A$ to a final one $\hat h(\tau)= \hat h_B$ in a time~$t\in [0,\tau]$, where at the beginning and the end of the protocol the copies are non-interacting. This translates to the following boundary conditions
\begin{equation}
    \hat H_{\rm int} (0) = \hat H_{\rm int} (\tau) = 0~,\quad
    \hat h^{(j)}(0) = \hat h_A~,\quad \hat h^{(j)}(\tau) = \hat h_B\quad\forall j~.
\end{equation}
Since we are considering only protocols in which the initial and final Hamiltonian commute (cf. main text), the driving will not require us to change the eigenbasis of $\hat H(t)$. Therefore we can reduce the number of needed parameters to $n=d^N$ parameters to describe $\hat H(t)$ by choosing $\{\gamma^j(t)\}$ to be its eigenvalues at time $t$, and $\hat X_j$ to be the projector associated to $\gamma^j(t)$. The average work cost of the operation can be computed by: 
\begin{equation}
    W= \int_0^\tau\!\!dt~ \Tr\!\left[\frac{d\hat H(t)}{dt} \hat\rho(t)\right]~,
\end{equation}
where $\hat\rho(t)$ is the state of the system. We can split this expression into $W = \Delta F + W_{\rm diss}$, where $\Delta F= \beta^{-1} \ln \mathcal{Z}(0)/ \mathcal{Z}(\tau)$ (with $\mathcal{Z}(t)= \Tr\big[e^{-\beta \hat H(t)}\big]$) is the reversible contribution and depends only on the endpoints. Whereas the dissipative term $W_{\rm diss}$ depends on the specific driving $\hat H(t)$, at first order in the slow driving regime, it is quantified by
\begin{equation}
   W_{\rm diss}= k_B T\int_0^\tau dt~d{\gamma}^i(t)d{\gamma}^j(t) g_{ij}(\gamma(t)) + \mathcal{O}\!\left(\tau^{-2}\right) ~,
\end{equation}
where $g_{ij}(\gamma)$ is a metric over the manifold $M$ of the control parameters $\{\gamma^i\}$. This metric allows us to quantify the length of a line element $ds$ over the manifold $M$: $ds^2 = g_{ij}d\gamma^i d\gamma^j$. By integrating $ds$ over a curve $\gamma(t)$ we can find the length of the curve:
\begin{equation}
   L[\gamma] = \int_\gamma \!ds = \int_0^\tau \!dt \sqrt{ d{\gamma}^i(t)d{\gamma}^j(t) g_{ij}(\gamma(t))}~.
\end{equation}
The length of the curve and the dissipation of the corresponding driving are related by a Cauchy-Schwartz inequality
\begin{equation}
    \beta W_{\rm diss} \geq \frac{1}{\tau}L^2~,
\end{equation}
where equality is satisfied whenever the integrand is constant. Since geodesics -- curves of minimal length connecting two points -- have a constant integrand for the length, they also minimize dissipation. Therefore we can find protocols that minimize the dissipation by solving the geodesic equations:
\begin{equation}\label{sm:geo_eq}
    \ddot\gamma^i(t) + \Gamma^i_{jk} \dot\gamma^j(t)\dot\gamma^k(t) = 0~,
\end{equation}
where $\Gamma^i_{jk}$ are the Christoffel symbols
\begin{equation}\label{sm:kartoffel}
    \Gamma^i_{jk} = \frac{1}{2}g^{il}\left(\frac{\partial g_{lk}}{\partial \gamma^j} + \frac{\partial g_{jl}}{\partial\gamma^k} - \frac{\partial g_{jk}}{\partial\gamma^l}\right)~,
\end{equation}
for $g^{il}$ the inverse of the metric.\\

In this work, we focus on a simple model of thermalization with a single time-scale $\tau_{\rm eq}$, which is described by the rate equation
\begin{equation}
    \frac{d}{dt}\hat\rho(t) = \frac{\hat\omega_\beta(t) -\hat\rho(t)}{\tau_{\rm eq}}~,
\end{equation}
where $\hat\omega_\beta(t) = e^{-\beta\hat H(t)}/\mathcal Z(t)$ is the instantaneous thermal state. Without loss of generality, we can absorb the time scale in the total time of the protocol, which allows us to drop it for simplicity. This model guarantees us the following form for the metric~\cite{Abiuso2020Geometric}
\begin{equation}\label{sm:metric_Z}
    g_{ij} = \frac{\partial^2 \ln \mathcal{Z} }{\partial \gamma^i \partial \gamma^j }~. 
\end{equation}
Given the full control assumed here, and choosing the parametrization to be such that $\{\gamma^i\}$ correspond to the eigenenergies of $\hat H$, this leads to
\begin{equation}\label{sm:metric_fc}
    \beta^{-2}g_{ij} = \omega_i \delta_{ij} - \omega_i\omega_j~,
\end{equation}
where $\omega_i = e^{-\beta \gamma_i}/\mathcal Z$ is the thermal probability of the eigenstate corresponding to the energy $\gamma^i$.

\subsection{Recovering the quantum Hellinger angle}
We  now show that the distance function induced by the metric \eref{sm:metric_fc} is the quantum Hellinger angle
\begin{equation}
    d(\gamma,\gamma') = 2 \arccos\Tr\left[\sqrt{\hat\omega_\beta(\gamma)} \sqrt{\hat\omega_\beta(\gamma')}\right]~,
\end{equation}
where $\gamma$ and $\gamma'$ are two points in $M$, $\hat\omega_\beta(\gamma)$ and $\hat\omega_\beta(\gamma')$ are the corresponding thermal states. Let us notice that we can rewrite the Hellinger angle as $d(\gamma,\gamma') = 2 \arccos\sqrt{\omega^i \omega_i'}$ since the thermal states have the same eigenbasis. \\

Let us now consider the radius 2 sphere embedded in $\mathbb R^n$, we can describe it with Cartesian coordinates $\{r^i\}_{i=1}^n$ subject to the constraint $r^i r_i = 4$ . With the Euclidean metric, the line element in $\mathbb R^n$ is $dl^2 = \delta_{ij}dr^idr^j$, this naturally induces the notion of Euclidean distance over $\mathbb R^n$. By restricting ourselves to the radius 2 sphere it is clear that the Euclidean distance between a point $r$ and $r'$ on the sphere is given by the angle between these two points times the radius, which gives us 
\begin{equation}
\label{eq:fisher_thermal}
    d(r,r') = 2\arccos \frac{r^i r_i'}{4}~,
\end{equation}
where $r^i r_i'$ is the scalar product of $r$ and $r'$. We can notice that if we identify $r^i =2\sqrt{\omega^i}$ we recover the Hellinger angle, and the constraint is naturally satisfied by the thermal probabilities since it becomes $\sum_i\omega^i = 1$. Therefore by applying a coordinate transformation, we can recover the line element (or equivalently the metric) that induces the Hellinger angle as its distance in terms of variations of the eigenenergies instead of variations of the square root of the thermal probabilities. We start by transforming to the thermal probabilities as coordinates, we have 
\begin{equation}\label{sm:metric_fc_prob}
    dl^2 = \frac{\delta_{ij}}{\omega^i}d\omega^i d\omega^j~,
\end{equation}
where we used $dr^i = \frac{1}{\sqrt{\omega^i}}d\omega^i$. Using the definition of the thermal probabilities we can easily find $d\omega^i = \beta(\omega^i\omega^j - \delta^{ij}\omega^i)d\gamma_j$, with \eref{sm:metric_fc_prob} and the constraint $\sum_i \omega^i = 1$ we find $dl^2 = \beta^2(\omega_i \delta_{ij} - \omega_i\omega_j)d\gamma^id\gamma^j$, which corresponds to the metric we found in \eref{sm:metric_fc}. This reveals that the manifold $M$ is the positive quadrant of the $n$-dimensional sphere of radius 2. It confirms that the thermodynamic length is given by the Hellinger angle and allows us to find the corresponding optimal protocols over the space of parameters.

\subsection{Geodesics}
For this subsection, we will take $\tau = 1$ for lightness of notation, as the results can be trivially generalized to $\tau \neq 1$. Since $M$ is the positive quadrant of the $n$-dimensional sphere of radius 2 equipped with the Euclidean metric, it is very simple to find the geodesics. In terms of the coordinates $\{r^i\}_i$ the geodesic connecting $r(0)$ to $r(1)$ is given by
\begin{equation}\label{sm:geo_r}
    r(t) = 2\frac{(1-u(t))r(0) + u(t)r(1)}{\|(1-u(t))r(0) + u(t)r(1)\|}~,
\end{equation}
where $u: [0,1]\rightarrow [0,1]$ is a bijective and increasing function that we choose such that $\|\dot r(t)\|$ is a constant, as this is equivalent to $\frac{ds}{dt}$ being constant. This trajectory follows a great circle of the $n$-dimensional sphere. To compute $u(t)$ we can start by noticing that, because of the normalization, $\mathcal L = d(r(0),r(1)) = \frac{ds}{dt} = \dot r^i(t)\dot r_i(t)$ for a geodesic. Therefore we get the following differential equation for $u(t)$
\begin{equation}
    \mathcal L = \frac{2 \dot u(t) \sin\!\frac{\mathcal L}{2}}{1-2u(t)(1-u(t))(1-\cos\!\frac{\mathcal L}{2})}~,
\end{equation}
which can be solved by integration. We thus find
\begin{equation}
    u(t) = \frac{1}{2}\left(1+\frac{\tan\!\left[\frac{\mathcal L}{4}(2t-1)\right]}{\tan\!\frac{\mathcal L}{4}\!}\right)~.
\end{equation}
Plugging this result in \eref{sm:geo_r} and transforming the coordinates we find the geodesic in terms of the thermal state
\begin{equation}
    \hat \omega_\beta(t) = \frac{1}{\sin^2\!\frac{\mathcal L}{2}} \left(\sin\!\left[\frac{\mathcal L}{2}(1-t)\right] \sqrt{\hat\omega_\beta(0)} + \sin\!\left[\frac{\mathcal L}{2}t\right] \sqrt{\hat\omega_\beta(1)}\right)^2 ~.
\end{equation}
At this point, it is immediate that the geodesic trajectory for the Hamiltonian is
\begin{equation}\label{sm:geo_H}
    \hat H(t) \!=\! -2k_B T\log\!\left[ \sin\!\left[\frac{\mathcal L(\tau-t)}{2\tau}\right] \frac{e^{-\frac{\beta}{2}\hat H(0)}}{\mathcal Z(0)^{\frac{1}{2}} }  + \sin\!\left[\frac{\mathcal Lt}{2\tau}\right] \frac{e^{-\frac{\beta}{2}\hat H(\tau)}}{\mathcal Z(\tau)^{\frac{1}{2}} } \right]~,
\end{equation}
where we neglected terms proportional to the identity. The geodesic of \eref{sm:geo_H} describes the optimal trajectory that every energy level should follow, one can notice that if two distinct energy levels have the same boundary conditions then they follow the same trajectory. Therefore, by permutation invariance, the number of distinct trajectories is given by the number of different (i.e. without counting the degeneracies) energy levels in the initial and final Hamiltonian. By taking into account that the initial and final Hamiltonian do not have interaction terms and are permutation invariant we can conclude that there are at most only $n = \binom{N+d-1}{N} = \mathcal O(N^{d-1})$ distinct control parameters instead of $d^N$.\\

Let us now consider the question of what orders of interaction are present in the trajectory described by \eref{sm:geo_H}. The following argument is made for spins-$\frac{1}{2}$, but its generalization is immediate. Let us suppose that there is no $k$-th order interaction term (i.e. that it involves $k$ sites) in the trajectory in \eref{sm:geo_H}, then we denote by $\eps_k(t)$ an eigen-energy of $\hat H$ that corresponds to $k$ distinct excitations at time $t$. Then $\eps_k(t)$ can be written as a linear combination of eigen-energies corresponding to fewer excitations: $\eps_k(t) = \sum_{j=1}^{k-1}\sum_{l}^{N_j}\alpha_{j,l}\eps_{j,l}(t)$, where $\eps_{j,l}(t)$ are the eigen-energies corresponding to $j$ distinct excitations (we are supposing w.l.o.g. that there are $N_j$ of them) and $\alpha_{j,l}$ are the real numbers composing the linear combination. Inserting this into \eref{sm:geo_H} we get that the following must be satisfied for all $t\in[0,\tau]$
\begin{equation}
    \sum_{j,l}\alpha_{j,l}\log\!\left[A(t) e^{-\beta\eps_{j,l}(0)/2} + B(t) e^{-\beta\eps_{j,l}(\tau)/2}\right] = \log\!\left[A(t) e^{-\beta\sum_{j,l}\alpha_{i,j}\eps_{j,l}(0)/2} + B(t) e^{-\beta\sum_{j,l}\alpha_{i,j}\eps_{j,l}(\tau)/2}\right]~.
\end{equation}

By defining $\Delta \eps_{j,l} := \eps_{j,l}(\tau) - \eps_{j,l}(0)$ we can turn this last equation into
\begin{equation}\label{eq:contr}
    \sum_{j,l}\alpha_{j,l}\log\!\left[A(t) + B(t) e^{-\beta\Delta\eps_{j,l}/2}\right] = \log\!\left[A(t)+ B(t) e^{-\beta\sum_{j,l}\alpha_{j,l}\Delta\eps_{j,l}/2}\right]~.
\end{equation}
Since the function $\log\left(A + B \exp[\cdot]\right)$ is non-linear, this equality generally cannot be satisfied for all times for a given set of boundary conditions $\{\eps_{j,l}(0),\eps_{j,l}(\tau)\}_{j,l}$. It is worth noting that there are some examples in which $\eps_k(t)$ is diverging for all times where this equality will be satisfied and $k$-th order interactions are not needed. Furthermore, let us point out that permutation invariance is not necessary here, we only need to require that $\hat h^{(j)}(0) \neq \hat h^{(j)}(\tau)$ for all $j$. Therefore, generally \eref{eq:contr} cannot be satisfied. The only way to solve the contradiction is by removing the assumption that the $k$-th order terms are missing. We conclude that, generally, all orders of interactions (from $2$-body to $N$-body) are necessary to realize the geodesic trajectory.

\section{Fisher metric and minimum Hellinger angle on conditional probabilities}
\label{B}
We analyze here the metric~\eref{sm:metric_fc_prob}, also known as Fisher metric, for generic probability distributions having 1-way conditional dependence.
That is, consider a probability vector of the form
\begin{align}
    p_{i_1,i_2,i_3,\dots,i_m}=p^{(1)}_{i_1} p^{(2)}_{i_2|i_1} p^{(3)}_{i_3|i_2}\dots p^{(m)}_{i_m|i_{m-1}}\;,
    \label{eq:p_cond}
\end{align}
where $p^{(l)}_{i_l|i_{l-1}}$ is the conditional probability of outcome $i_l$ given the value of $i_{l-1}$.
Moreover, we will use the intuitive notation for marginal probabilities, that is, for example
\begin{align}
    p_{i_2,i_5}:=\sum_{i_1,i_3,i_4,i_6,\dots,i_m} p_{i_1,i_2,i_3,\dots,i_m}\;.
\end{align}
Notice in particular that with this notation $p^{(1)}_{i_1}\equiv p_{i_1}$.
For conditioned probabilities of the kind above~\eref{eq:p_cond}, the Fisher metric takes a special decomposition, that is
\begin{align}
\label{eq:Fisher-Markov}
    \sum_{\vec{i}} \frac{d p_{\vec i}^2}{p_{\vec i}}=\sum_{i_1} \frac{d p_{i_1}^2}{p_{i_1}} 
    + \sum_{i_1,i_2} p_{i_1}\frac{d p_{i_2|i_1}^{_{(2)}2}}{p_{i_2|i_1}}
    + \sum_{i_2,i_3} p_{i_2}\frac{d p_{i_3|i_2}^{_{(3)}2}}{p_{i_3|i_2}}
    +\dots
    +\sum_{i_{m-1},i_m} p_{i_{m-1}}\frac{d p_{i_m|i_{m-1}}^{_{(m)}2}}{p_{i_m|i_{m-1}}}\;.
\end{align}
This equation will be our central observation in the following, and can be derived explicitly.

\paragraph*{Proof of~\eref{eq:Fisher-Markov}.} We show here the case $m=3$, which can be generalized trivially. One has
\begin{align}
    \sum_{i_1,i_2,i_3}\frac{d p_{i_1,i_2,i_3}^{2}}{p_{i_1,i_2,i_3}}=\sum_{i_1,i_2,i_3}\frac{\left( d p^{(1)}_{i_1}p^{(2)}_{i_2|i_1}p^{(3)}_{i_3|i_2} + p^{(1)}_{i_1}d p^{(2)}_{i_2|i_1}p^{(3)}_{i_3|i_2} + p^{(1)}_{i_1}p^{(2)}_{i_2|i_1} d p^{(3)}_{i_3|i_2}\right)^{2}}{p^{(1)}_{i_1}p^{(2)}_{i_2|i_1}p^{(3)}_{i_3|i_2}}\;,
\end{align}
from which the numerator yields
\begin{multline}
    \sum_{i_1,i_2,i_3}\frac{d p^{_{(1)}2}_{i_1}p^{_{(2)}2}_{i_2|i_1}p^{_{(3)}2}_{i_3|i_2} + p^{_{(1)}2}_{i_1}d p^{_{(2)}2}_{i_2|i_1}p^{_{(3)}2}_{i_3|i_2} + p^{_{(1)}2}_{i_1}p^{_{(2)}2}_{i_2|i_1} d p^{_{(3)}2}_{i_3|i_2}}{p^{_{(1)}}_{i_1}p^{_{(2)}}_{i_2|i_1}p^{_{(3)}}_{i_3|i_2}} + \text{cross-terms}\\
    =\sum_{i_1,i_2,i_3}\frac{d p^{_{(1)}2}_{i_1}}{p^{_{(1)}}_{i_1}}p^{_{(2)}}_{i_2|i_1}p^{_{(3)}}_{i_3|i_2} + p^{_{(1)}}_{i_1}\frac{d p^{_{(2)}2}_{i_2|i_1}}{p^{_{(2)}}_{i_2|i_1}}p^{_{(3)}}_{i_3|i_2} + p^{_{(1)}}_{i_1}p^{_{(2)}}_{i_2|i_1} \frac{d p^{_{(3)}2}_{i_3|i_2}}{p^{_{(3)}}_{i_3|i_2}} + \text{cross-terms}\\
    =\sum_{i_1} \frac{d p^{2}_{i_1}}{p_{i_1}} + \sum_{i_1,i_2} p_{i_1}\frac{d p^{_{(2)}2}_{i_2|i_1}}{p^{_{(2)}}_{i_2|i_1}} + \sum_{i_2,i_3} p_{i_2} \frac{d p^{_{(3)}2}_{i_3|i_2}}{p^{_{(3)}}_{i_3|i_2}} + \sum_{i_1,i_2,i_3}\text{cross-terms}\;,
\end{multline}
where we simplified the expression by partially summing on the indices, using the identification $p^{(1)}_{i_1}\equiv p_{i_1}$, noticing that $\sum_{i_1}p_{i_1}p^{(2)}_{i_2|i_1}=p_{i_2}$, and using the normalization constraints on conditional probabilities.\\
To conclude the proof of~\eref{eq:Fisher-Markov} one needs to show that the cross-terms are null. This is indeed the case as
\begin{align}
    \sum_{i_1,i_2,i_3} d{p}^{(1)}_{i_1}d{p}^{(2)}_{i_2|i_1}{p}^{(3)}_{i_3|i_2} &= 0 \text{ by summing on } i_3 \text{ first, and then on } i_2\;,\\
    \sum_{i_1,i_2,i_3} d{p}^{(1)}_{i_1}{p}^{(2)}_{i_2|i_1}d{p}^{(3)}_{i_3|i_2} &= 0 \text{ by summing on } i_3\;,\\
    \sum_{i_1,i_2,i_3} {p}^{(1)}_{i_1}d{p}^{(2)}_{i_2|i_1}d{p}^{(3)}_{i_3|i_2} &= 0 \text{ by summing on } i_3\;.
\end{align}
This concludes the proof.

\subsection{Bounds on the Fisher distance between conditional probabilities}
In this section, we prove a simple bound on the geodesic Fisher distance between probabilities with conditional dependence, for a specific form of~\eref{eq:Fisher-Markov}.
Namely, consider a probability vector of the form
\begin{align}
\label{eq:p_cond_3}
    p_{i_1, i_2, i_3}=p^{(1)}_{i_1} p_{i_2|i_1}^{(2)} p_{i_3}^{(3)}\;.
\end{align}
In the following we will assume \emph{full control} on $p^{(1)}_{i_1}$, while $p^{(2)}_{i_2|i_1}$ might be constrained to belong to a submanifold of conditional probabilities; finally the index $i_3$ includes all the degrees of freedom that are uncorrelated to $p_{i_1,i_2}$. Moreover, our results will be general to any cardinality of the various indexes.
As we are interested in Landauer erasures, and as a tool for simplification, we consider the distance to partially deterministic distributions. That is, probability vectors representing a deterministic outcome for $i_1$ and $i_2$, which without loss of generality can be expressed (via re-indexing) always as
\begin{align}
    p^{\rm det}_{i_1,i_2}=\delta_{i_1,0}\delta_{i_2,0}\;.
    \label{eq:p_det}
\end{align}
where $\delta_{i,j}$ is the Kronecker product. Notice that $p^{\rm det}$ is a particular case of the conditional form \eref{eq:p_cond}.

We can now state our main Lemma
\begin{lem}
\label{lem:cond_m}
The geodesic Fisher distance between any $p$ of the form \eref{eq:p_cond_3}, when fixing $p^{(3)}$ and assuming full control on $p^{(1)}$, is bounded by $3\pi$. When the final point is deterministic on $i_1$ and $i_2$, the bound can be reduced to $2\pi$.
That is, for fixed boundary conditions
\begin{align}
\label{eq:boundary_m3}
    p_{\vec{i}}(0)\equiv p^{(1)}_{i_1}(0)\, p^{{(2)}}_{i_2|i_1}\!(0)\, p_{i_3}^{(3)}\;, \quad {\rm and} \quad p_{\vec{i}}(1)\equiv p^{(1)}_{i_1}(1)\, p_{i_2|i_1}^{(2)}\!(1)\, p_{i_3}^{(3)}\;,
\end{align}
and assuming full control on $p^{(1)}$ only, one has
\begin{align}
\label{eq:lemma_ineq_gen}
   \min_{p_{\vec{i}}(t)\equiv p^{(1)}_{i_1}(t)\, p_{i_2|i_1}^{(2)}\!(t)\, p_{i_3}^{(3)} } \int_0^1 \!dt ~\frac{{\dot p}^2_{\vec{i}}}{p_{\vec{i}}} \; \leq 3\pi\;.
\end{align}
Moreover, if the final point is of the form \eref{eq:p_det}, the same bound reads
\begin{align}
\label{eq:lemma_ineq_lan}
   \min_{p_{\vec{i}}(t)\equiv p^{(1)}_{i_1}(t)\, p_{i_2|i_1}^{(2)}\!(t)\, p_{i_3}^{(3)} } \int_0^1 \!dt ~\frac{{\dot p}^2_{\vec{i}}}{p_{\vec{i}}} \; \leq 2\pi\;\quad {\rm when \quad} p_{\vec{i}}(1)\equiv \delta_{i_1,0}\delta_{i_2,0} p_{i_3}^{(3)}\;.
\end{align}
\end{lem}
As we are mainly interested in Landauer erasure, in the following applications we will mainly use the second inequality \eref{eq:lemma_ineq_lan}.
The proof of the Lemma is constructive. That is, we show an explicit trajectory $p(t)$ that satisfies \eref{eq:lemma_ineq_gen} and \eref{eq:lemma_ineq_lan}.

\paragraph*{Proof of \lemref{lem:cond_m}.}
To prove \lemref{lem:cond_m}, we consider a trajectory $p^{\rm step}(t)$ of the form~\eref{eq:p_cond_3} 
\begin{align}
    p^{\rm step}_{\vec{i}}(t)=p^{(1)}_{i_1}(t)\, p_{i_2|i_1}^{(2)}\!(t)\, p_{i_3}^{(3)} 
\end{align}
and use~\eref{eq:Fisher-Markov} specialized to such case as
\begin{align}
\label{eq:metric_cond_simp}
     \sum_{\vec{i}} \frac{dp_{\vec i}^2}{p_{\vec i}}=\sum_{i_1} \frac{ dp_{i_1}^2}{p_{i_1}} 
    + \sum_{i_1,i_2} p_{i_1}\frac{dp_{i_2|i_1}^{_{(2)}2}}{p_{i_2|i_1}}\;.
\end{align}

We consider $p^{\rm step}(t)$ to follow 5 steps. In steps $1$, $3$, and $5$, the distance traveled is bounded by $\pi$, while in steps $2$ and $4$ the distance is null. By the triangular inequality, the Lemma is then proven. In particular, if the final point is deterministic on $i_1$ and $i_2$, Step 5 is not needed and the $2\pi$ bound~\eref{eq:lemma_ineq_lan} follows. For more general endpoints Step 5 is needed, proving~\eref{eq:lemma_ineq_gen}.

\begin{itemize}
    \item \underline{Step $1$.} First $p_{i_1}^{(1)}(t)$ goes to a deterministic distribution, w.l.o.g.
    \begin{align}
        p^{(1)}_{i_1}(0)\rightarrow \delta_{i_1,1}
    \end{align}
    while $p^{(2)}$ and (by hypothesis) $p^{(3)}$ do not change. It follows from~\eref{eq:metric_cond_simp} that the distance consumed during this step is given only by $\int\sum_{i_1} dt \frac{\dot p_{i_1}^{_{(1)}2}}{p_{i_1}}$, and therefore by choosing the geodesic path between $p^{(1)}_{i_1}(0)$ and $\delta_{i_1,1}$, one travels a distance that is bounded by $d(p^{(1)}_{i_1},\delta_{i_1,1})\leq\pi$.

    \item \underline{Step $2$.}
    Secondly, $p^{(2)}_{i_2|i_1}$ is modified to its final value for all values of $i_1$ except $i_1=1$, 
    \begin{align}
        p^{(2)}_{i_2|i_1}\rightarrow p^{(2)}_{i_2|i_1}\!(1) \quad \forall i_1\neq 1 \;,
    \end{align}
    while all other probabilities are fixed. Notice that in the case of an erasure protocol, one has $p^{(2)}_{i_2|i_1}\!(1)\equiv \delta_{i_2,0} $. Due to $p^{(1)}_{i_1}=\delta_{i_1,1}$ and~\eref{eq:Fisher-Markov} it follows that the distance traveled by the whole probability distribution is null (as the weight associated with $i_1\neq 1$ is null).

    \item \underline{Step $3$.} As a third step, one modifies again $p^{(1)}_{i_1}$, bringing it to a different deterministic value \eref{eq:p_det}, i.e.
    \begin{align}
        p^{(1)}_{i_1}=\delta_{i_1,1}&\rightarrow \delta_{i_1,0}\;.
    \end{align}
    Once again, the distance cost of this step is bounded by $\pi$, when taking the geodesic path from $\delta_{i_1,1}$ to $\delta_{i_1,0}$.

    \item \underline{Step $4$.} The fourth step takes $p^{(2)}_{i_2|i_1=1}$ to its final value
    \begin{align}
        p^{(2)}_{i_2|1}&\rightarrow   p^{(2)}_{i_2|1}\!(1)
    \end{align}
    at zero distance cost, due to $p_{i_1=1}=0$. Notice that at this point the full vector $p^{(2)}_{i_2|i_1}\!(t)$ is in its final point and $p^{(1)}$ is in a deterministic configuration.

    \item \underline{Step $5$ (only for non-deterministic $p^{(1)}_{i_1}(1)$).} If the final desired point is non-deterministic in the $i_1$ index, a final step is needed to complete
    \begin{align}
        p^{(1)}_{i_1}=\delta_{i_1,0}\rightarrow p^{(1)}_{i_1}\!(1)\;.
    \end{align}
    This step consists again of traveling a distance smaller than $\pi$.
   
\end{itemize}
Summarizing, the total effect of the above steps is that of transforming
    \begin{align}
        p^{\rm step}(0)=p^{(1)}_{i_1} (0)p^{(2)}_{i_2|i_1}\!(0) p^{(3)}_{i_3}\rightarrow p^{\rm step}(1)=p^{(1)}_{i_1} (1)p^{(2)}_{i_2|i_1}\!(1) p^{(3)}_{i_3}
    \end{align}
    at a distance cost bounded by $2\pi$ for deterministic final points (erasure protocols), or $3\pi$ otherwise. This proves \lemref{lem:cond_m}.

\subsection{Star Model}
\begin{figure}
    \centering
    \includegraphics[width=0.3\textwidth]{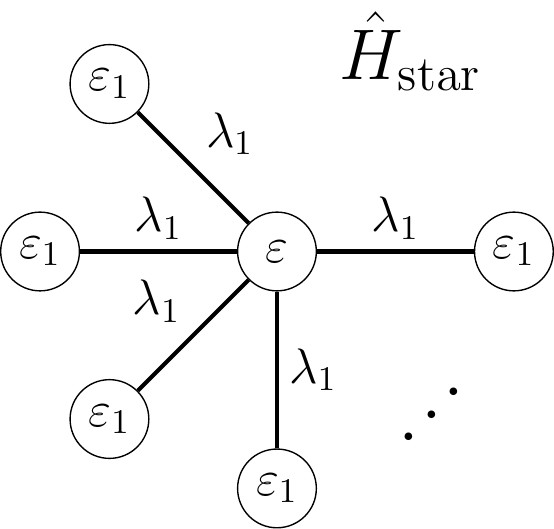}
    \caption{Star model: a central spin }
    \label{fig:star}
\end{figure}

Consider the following Star-shaped Hamiltonian, featuring a central spin interacting with all the remaining ones, as depicted in \fref{fig:star}
\begin{align}
\label{eq:H_star}
    \hat H_{\rm star}(t) &= \eps(t)\hat\sigma_z^{(1)} + \lambda_1(t)\sum_{i=2}^N \hat\sigma_z^{(i)} + \lambda_2(t)\sum_{i=2}^N \hat\sigma_z^{(1)}\hat\sigma_z^{(i)} \;.
\end{align}
This particular choice of interactions creates on the $N-1$ external spins an effective uniform magnetic field that is equal to $\lambda_1+\lambda_2$ when $\hat{\sigma}_z^{(1)}=+1$, while $\lambda_1-\lambda_2$ when $\hat{\sigma}_z^{(1)}=-1$. For this reason, the change of variable
\begin{align}
\lambda_+=\lambda_1+\lambda_2\;, \quad \lambda_-=\lambda_1-\lambda_2\;,    
\end{align}
makes the computation of the thermal probabilities and the partition function particularly easy.
In particular, choosing temperature units in which $\beta=1$,
\begin{align}
    \mathcal{Z}_{\rm star}=e^{-\eps}(2\cosh{\lambda_+})^{N-1}+e^{\eps}(2\cosh{\lambda_-})^{N-1}\;.
\end{align}
Moreover, one can notice how the Hamiltonian \eref{eq:H_star} leads to a thermal state $\omega_{i_1,i_2}$ of the form \eref{eq:p_cond} for $m=2$. In this case, $i_1=\pm 1$ represents the two possible configurations of the central spin $\hat{\sigma}^{(1)}_z$, while $\vec{i_2}=\{+1,-1\}^{N-1}$ labels the configuration of the remaining $N-1$ spins.
It is straightforward to compute the probabilities of the central spin
\begin{align}
    \omega_{i_1=+1}&=p^{(1)}_{i_1=+1}=\frac{e^{-\eps}(2\cosh{\lambda_+})^{N-1}}{e^{-\eps}(2\cosh{\lambda_+})^{N-1}+e^{\eps}(2\cosh{\lambda_-})^{N-1}}\;,\\
    \omega_{i_1=-1}&=p^{(1)}_{i_1=-1}=\frac{e^{\eps}(2\cosh{\lambda_-})^{N-1}}{e^{-\eps}(2\cosh{\lambda_+})^{N-1}+e^{\eps}(2\cosh{\lambda_-})^{N-1}}\;.
\end{align}
Similarly, once the central spin is fixed i.e. $\hat{\sigma}^{(1)}_z=i_1$, the statistics of the remaining spins are defined by a homogeneous local magnetic field equal to $\lambda_{i_1}$,
\begin{align}
    \omega_{\vec{i_2}|i_1}=p^{(2)}_{\vec{i_2}|i_1}=\frac{e^{-\lambda_{i_1}\sum_{i=2}^{N}\sigma_z^{(i)}}}{(2\cosh{\lambda_{i_1}})^{N-1}}\;,
\end{align}
which can be seen to factorize in parallel, that is
\begin{align}
  \omega_{\vec{i_2}|i_1}=  \omega_{i_2|i_1}^{\otimes N-1}\;.
\end{align}
The resulting metric becomes of the form
\begin{align}
    \sum_{i_1} \frac{d\omega_{i_1}^2}{\omega_{i_1}} 
    + \sum_{i_1,\vec{i_2}} \omega_{i_1}\frac{d\omega_{\vec{i_2}|i_1}^{2}}{\omega_{\vec{i_2}|i_1}}=
    \sum_{i_1} \frac{d\omega_{i_1}^2}{\omega_{i_1}} 
    +(N-1) \sum_{i_1,{i_2}} \omega_{i_1}\frac{d\omega_{i_2|i_1}^{2}}{\omega_{i_2|i_1}}\;.
\end{align}

The above discussion clarifies how the Hamiltonian control~\eref{eq:H_star} induces a control on the thermal statistics $\hat \omega$ which is of the form~\eref{eq:p_cond_3}.
To bound the dissipation for such a model, we can therefore invoke \lemref{lem:cond_m}, which provides the existence of an erasure protocol, starting from any initial thermal state $\hat \omega$ to a final deterministic $\hat \omega^{\rm det}$, with Fisher distance bounded by bound
\begin{align}
    L^{\rm Star}(\omega,\omega^{\rm det})\leq 2\pi\;,
\end{align}
which is translated to an upper bound on the dissipation
\begin{align}
   \tau \beta W_{\rm diss}\equiv L^{\rm Star}(\omega,\omega^{\rm det})^2\leq 4\pi^2\;.
\end{align}

\subsection{Pyramid scheme}
\begin{figure}
    \centering
    \includegraphics[width=0.6\textwidth]{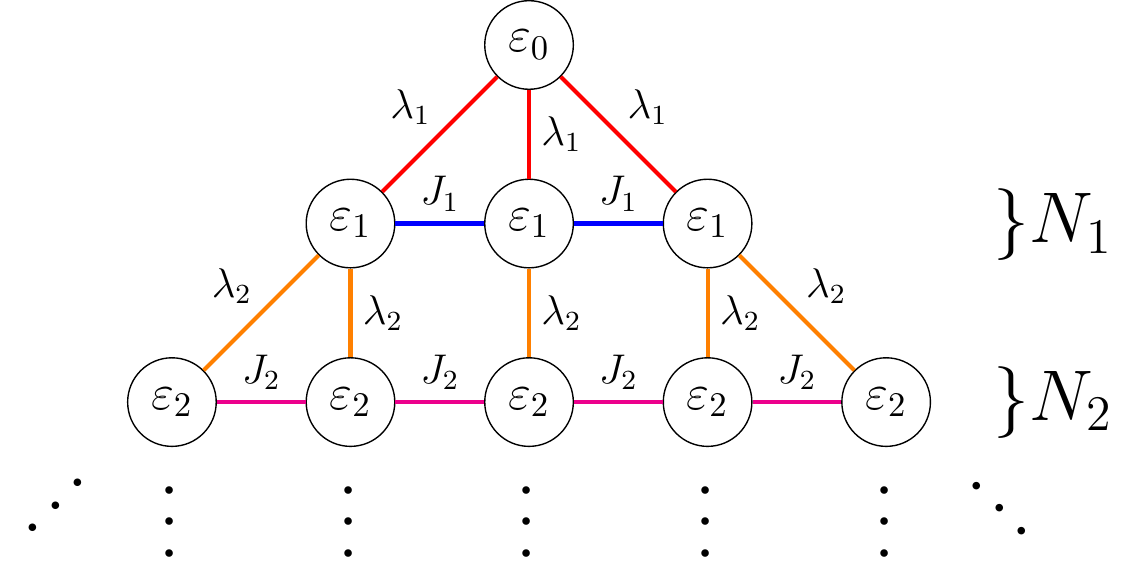}
    \caption{Pyramid model with $N_i=1+2i$ (cf.~\eref{eq:Npyr_mod}), in dimension $D=2$.}
    \label{fig:cone}
\end{figure}

In this section, we generalize the Star model above and its minimally-dissipating protocol to a multi-layer structure, which importantly features only \emph{short-range} interactions. Consider the pyramidal structure presented in \fref{fig:cone}, which can be ascribed to a Hamiltonian of the form
\begin{multline}
\label{eq:H_cone}
    \hat H_{\rm pyr}(t) =
    \eps(t)\hat\sigma_z^{(1)}
    + \eps_1(t)\sum_{i_1=1}^{N_1} \hat\sigma_z^{(i_1)}
    + \eps_2(t)\sum_{i_1=1}^{N_2} \hat\sigma_z^{(i_2)} + \dots
    +\eps_{m-1}(t)\sum_{i_{m-1}=1}^{N_{m-1}} \hat\sigma_z^{(i_{m-1})}
    \\
    + \lambda_{1}(t)\sum_{i_1=1}^{N_1} \hat\sigma_z^{(1)}\hat\sigma_z^{(i_1)} 
    + \lambda_{2}(t)\sum_{i_1,i_2=1,1}^{N_1,N_2} \hat\sigma_z^{(i_1)}\hat\sigma_z^{(i_2)} +
    \dots + \lambda_{m-1}\sum_{i_{m-2},i_{m-1}=1,1}^{N_{m-2},N_{m-1}}\hat\sigma_z^{(i_{m-2})}\hat\sigma_z^{(i_{m-1})}
    \\
    +J_1(t)\sum_{i_1=1}^{N_1}\hat{\sigma}_z^{(i_1)}\hat{\sigma}_z^{(i_1+1)}
    +J_2(t)\sum_{i_2=1}^{N_2}\hat{\sigma}_z^{(i_2)}\hat{\sigma}_z^{(i_2+1)}+\dots
    +J_{m-1}(t)\sum_{i_{m-1}=1}^{N_{m-1}}\hat{\sigma}_z^{(i_{m-1})}\hat{\sigma}_z^{(i_{m-1}+1)}
    \;.
\end{multline}

We now consider an erasure protocol for the such pyramid model, starting from a completely uncorrelated thermal state of the form 
\begin{align}
\omega_{i_0,i_1,i_2,\dots,i_{m-1}}^{\rm initial}=\omega_{i_0}\omega_{i_1}\omega_{i_2}\dots\omega_{i_{m-1}}\;,
\end{align}
corresponding to all $\lambda_i=0$ and $J_i=0$ in the Hamiltonian~\eref{eq:H_cone},
to a final deterministic configuration
\begin{align}
\omega_{i_0,i_1,i_2,\dots,i_{m-1}}^{\rm final}=\delta_{i_0,0}\delta_{i_1,0}\delta_{i_2,0}\dots\delta_{i_{m-1},0}\;.
\end{align}
To bound the total dissipation, we consider a step protocol of the form
\begin{multline}
    \omega_{i_0}\omega_{i_1}\omega_{i_2}\dots\omega_{i_{m-1}} 
    \rightarrow \delta_{i_0,0}\delta_{i_1,0}\omega_{i_2}\dots\omega_{i_{m-1}}  
    \rightarrow \delta_{i_0,0}\delta_{i_1,0}\delta_{i_2,0}\dots\omega_{i_{m-1}} 
    \rightarrow \dots 
    \rightarrow \delta_{i_0,0}\delta_{i_1,0}\delta_{i_2,0}\dots\delta_{i_{m-1},0}\;.
\end{multline}
That is, at each step $k$, an erasure is completed for the $(k-1)$-th and $k$-th layer, while the other layers are left untouched, for a total of $m-1$ steps.
We now claim that each of the $m-1$ steps fulfills the hypothesis of \lemref{lem:cond_m} when appropriately using the controls of the Pyramid Hamiltonian~\eref{eq:H_cone}.
In fact, each $k$-step has boundary conditions of the form~\eref{eq:boundary_m3}, by identifying $p^{(1)}_{i_1}\leftrightarrow \omega^{\rm initial/final}_{i_{k-1}}$,  $p^{(2)}_{i_2|i_1}\leftrightarrow \omega^{\rm initial/final}_{i_{k}}$, and all the untouched degrees of freedom $i_3\leftrightarrow i_0,\dots,i_{k-2},i_{k+1},\dots,i_{m-1}$.
For the Lemma to be valid at each step, one needs to assume full-control on the $i_{k-1}$ degrees of freedom. This is not the case in general, as each layer corresponds to a uniform Ising-like control. However, this is resolved with the following care:
after step $k-1$, layer $k-1$ is in a deterministic $\omega_{i_{k-1}}=\delta_{i_{k-1},0}$. In such case, it is possible to modify the couplings $J_{k-1}\rightarrow\infty$ to a fully ferromagnetic configuration without modifying $\hat\omega$. This produces an effective two-level system, i.e. all the spins up or all spins down at layer $k-1$, on which the local magnetic fields act as an effective full-control for such a two-level system.

\begin{figure}[H]
    \centering
    \includegraphics[width=.8\columnwidth]{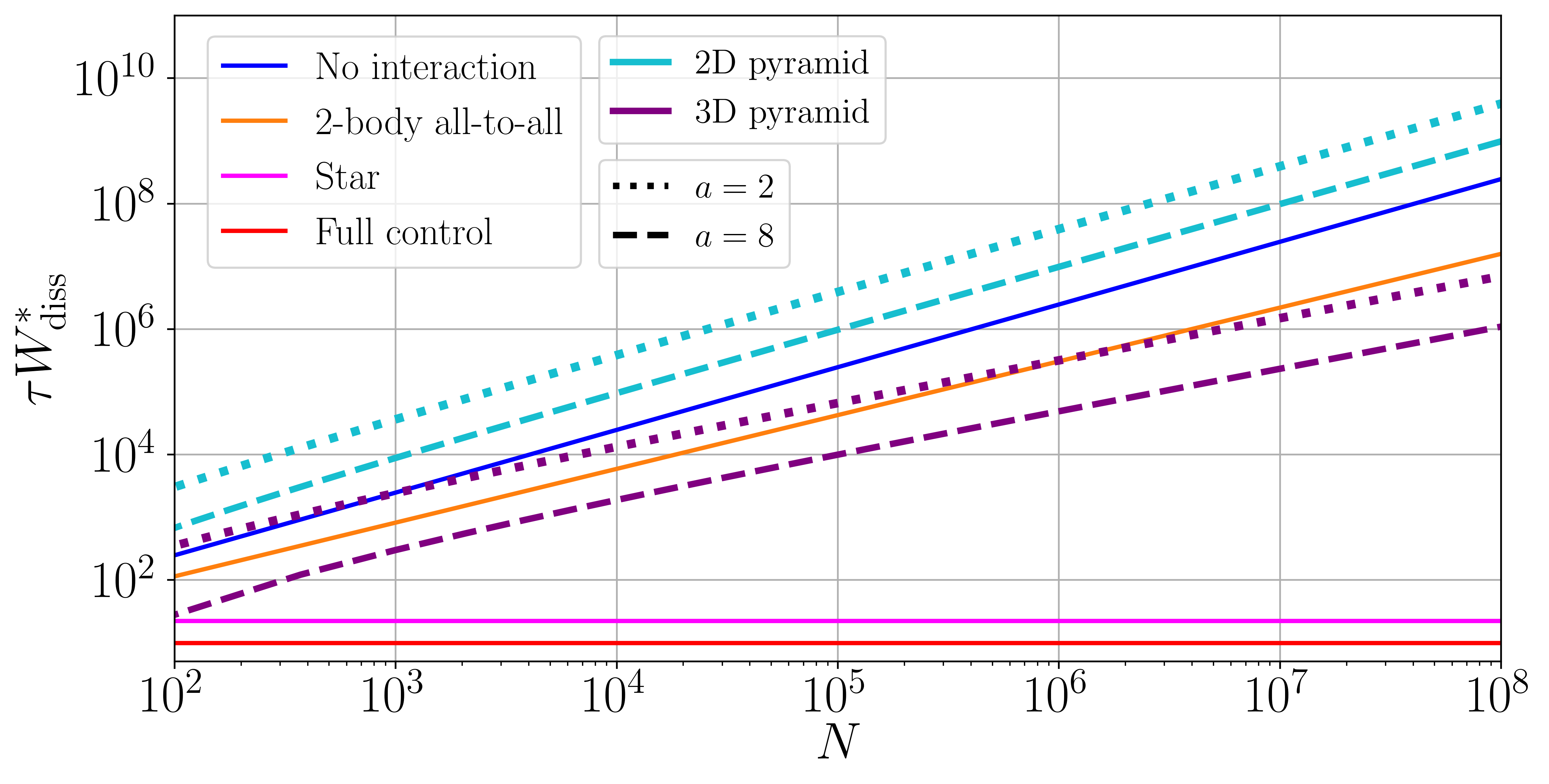}
    \vspace{-15pt}
    \caption{Minimal dissipation for the erasure of $N$ spins in the non-interacting case and full control scenario. The dissipation of the protocol described in \lemref{lem:cond_m} for the erasure of $N$ spins in the star model, and its repeated application for pyramid models in 2-D and 3-D with apertures $a = 2$ and $a=8$. And an extrapolation of the fit of the minimal dissipation in the all-to-all model (to compare the scaling) $W_{\rm diss}^{\rm *, all} = \alpha N^x$ with $x = 0.857$ and $\alpha = 2.20$.}
    \vspace{-10pt}
    \label{fig:pyramids_scalings}
\end{figure}

Given the above discussion, we can bound the dissipation in the Pyramid model using the $(m-1)$-step protocol therein described, which corresponds to a total Fisher distance $L\leq 2(m-1)\pi$. It then follows that the dissipations are bounded by 
\begin{align}
\label{eq:bound_layer}
    \tau \beta W_{\rm diss}\equiv L^2 \leq 4(m-1)^2\pi^2\;,
\end{align}
while the total number of spins is
\begin{align}
    N=N_0+\dots +N_{m-1}+N_{m-1}\;.
\end{align}
In standard pyramids with spatial dimension $D$, we can assume that
\begin{align}
\label{eq:Npyr_mod}
    N_i=(c+a (i-1))^{D-1}
\end{align}
for some positive integers $c$ and $a$. This corresponds to a pyramid-shaped Hamiltonian in spatial dimension $D$ having the first layer increase with size $a$. For example, the pyramid in ~\fref{fig:cone} corresponds to $c=1$ and $a=2$.\\
In such a model the total number of spins scales as
\begin{align}
    N=\sum_{i=0}^{m-1} N_i=\sum_{i=0}^{m-1} (1+a i)^{D-1}\approx \frac{a^{D-1}}{D}m^D
\end{align}
at leading order.\\
It follows that the dissipation~\eref{eq:bound_layer} can be expressed, at leading order,
\begin{equation}
    \tau \beta W_{\rm diss}\approx \frac{4\pi^2}{a^2}(a D N)^{\frac{2}{D}}\;.
\end{equation}
The above expression shows that the scaling of the dissipation in such models is sub-linear in $N$ as soon as $D>2$, (e.g. if the pyramid is 3-dimensional $D=3$). The obtained dissipation for $D=2$ and $D=3$ is showcased in \fref{fig:pyramids_scalings} where it's contrasted to the minimal dissipation found for the other models. 

\section{Numerics for many-body systems}
\label{C}
In this section, we explain the techniques used to find the protocols that minimize dissipation for the two-body all-to-all system of spins and the 1-D Ising chain. The Hamiltonians of these systems are the following
\begin{align}
    \hat H_{\rm all}(t) &= \eps(t)\sum_{i=1}^N\hat\sigma_z^{(i)} + \frac{1}{2}J(t)\sum_{i,j=1}^N \hat\sigma_z^{(i)}\hat\sigma_z^{(j)}~,\\
    \hat H_{\rm nn}(t) &= \eps(t)\sum_{i=1}^N\hat\sigma_z^{(i)} + \frac{1}{2}J(t)\sum_{i=1}^N \hat\sigma_z^{(i)}\hat\sigma_z^{(i+1)}~.
\end{align}
\begin{figure}[ht]
    \centering
    \includegraphics[width=.95\columnwidth]{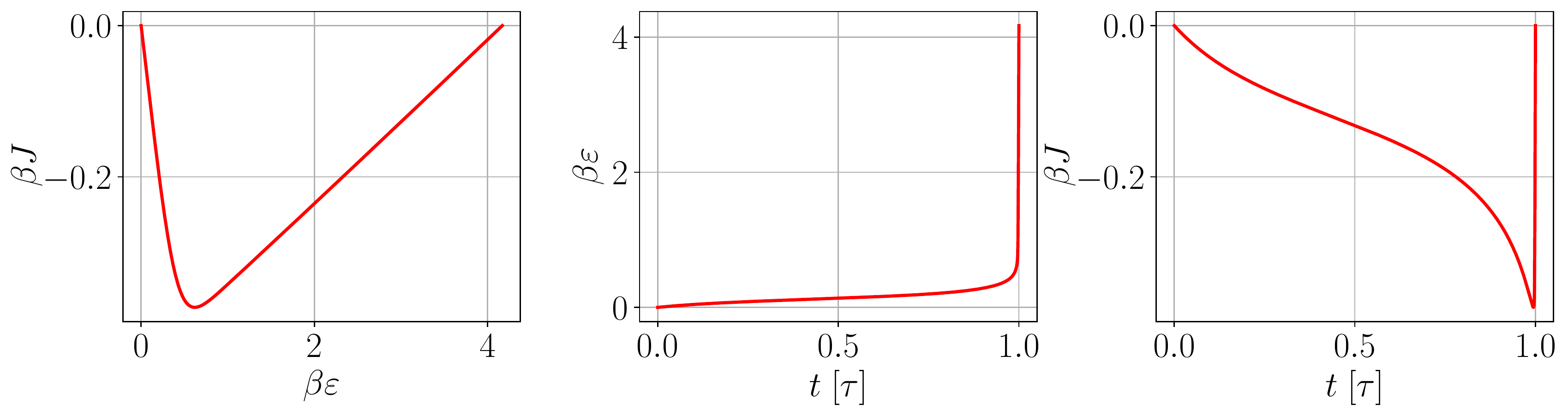}
    \vspace{-15pt}
    \caption{Optimal finite-time protocol for approximate erasure ($\eps(\tau) = 4 k_B T$ which has an erasure error of $3\cdot 10^{-4}$) in the all-to-all spin model. This protocol is computed for $N=10$.}
    \vspace{-10pt}
    \label{fig:geo_all}
\end{figure}

\begin{figure}[ht]
    \centering
    \includegraphics[width=.95\columnwidth]{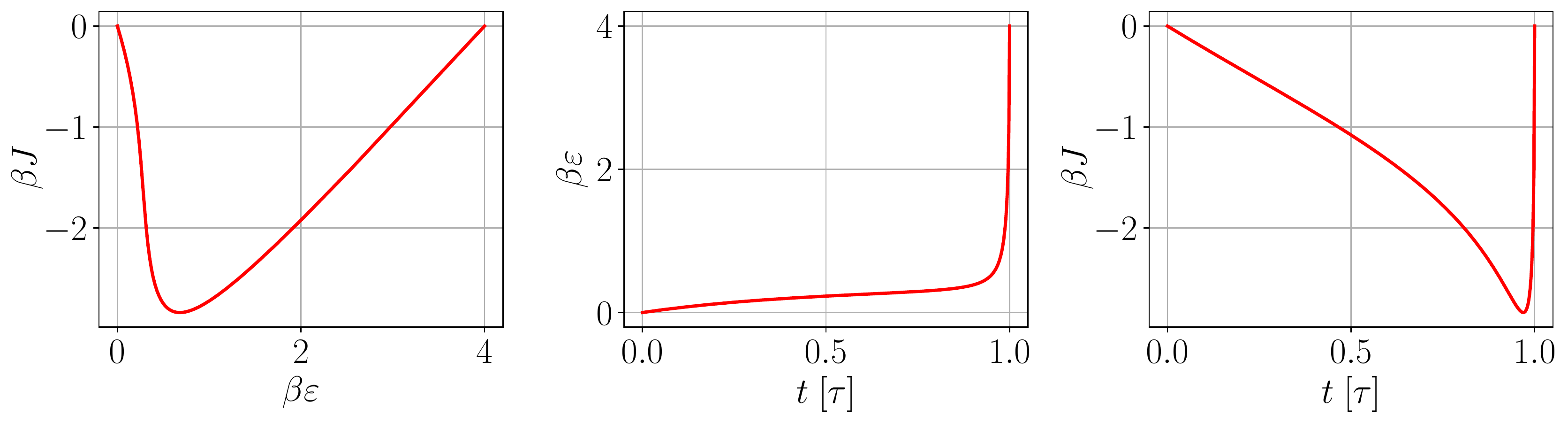}
    \vspace{-15pt}
    \caption{Optimal finite-time protocol for approximate erasure ($\eps(\tau) = 4 k_B T$ which has an erasure error of $3\cdot 10^{-4}$) in the 1-D Ising chain model. This protocol is the same for all values of $N$.}
    \vspace{-10pt}
    \label{fig:geo_1d}
\end{figure}
To find the optimal driving we use the formalism of geometric thermodynamics which explained in the \aref{A}. In particular, the optimal driving protocols solve the geodesic equation \eref{sm:geo_eq}. To compute it we need the Christoffel symbols, which can be computed from derivatives of the partition function \eref{sm:metric_Z}. In both scenarios at hand, the geodesic equation cannot be solved analytically. Therefore we want to be able to express the partition function (in particular its derivatives) in numerically tractable ways. For the all-to-all case, we can write the partition function and its derivatives as follows
\begin{align}
    \label{sm:all_Z}
    \mathcal Z_{\rm all} =&~ \sum_{k=0}^N \binom{N}{k} e^{-\beta E_k}~,\\
    \beta^{-2}\frac{\partial^2 \ln \mathcal{Z}_{\rm all} }{\partial \gamma^i \partial \gamma^j } =&~ \langle \hat X_i \hat X_j \rangle - \langle \hat X_i \rangle \langle \hat X_j \rangle~, \\
    -\beta^{-3}\frac{\partial^3 \ln \mathcal{Z}_{\rm all} }{\partial \gamma^i \partial \gamma^j \partial \gamma^k} =&~ \langle \hat X_i \hat X_j \hat X_k \rangle - \langle \hat X_i \hat X_j\rangle \langle \hat X_k \rangle - \langle \hat X_i \hat X_k\rangle \langle \hat X_j \rangle \\ & - \langle \hat X_k \hat X_j\rangle \langle \hat X_i \rangle + 2 \langle \hat X_i \rangle \langle \hat X_k \rangle \langle \hat X_k \rangle~,\nonumber
\end{align}
where we defined $E_k = \eps(2k-N) + \frac{1}{2}J(2k-N)^2$, $\gamma = (\eps, J)$, $\hat X_1 = \sum_{k=1}^N \hat \sigma_z^{(k)}$, $\hat X_2 = \frac{1}{2}\sum_{k,l=1}^N \hat \sigma_z^{(k)}\hat\sigma_z^{(l)}$, and the expectation values are computed with respect to the thermal state. Thanks to \eref{sm:all_Z} we can efficiently compute (for large $N$) these expectation values, as we get the following expression
\begin{equation}
    \langle \hat X_{i_1}\hat X_{i_2} ...\, \hat X_{i_j}\rangle = \frac{1}{\mathcal Z_{\rm all}}\sum_{k=0}^N \binom{N}{k} \frac{\partial E_k}{\partial \gamma^{i_1}} \frac{\partial E_k}{\partial \gamma^{i_2}}... \frac{\partial E_k}{\partial \gamma^{i_j}} e^{-\beta E_k}~.
\end{equation}
Whereas for the 1-D Ising chain we can compute the partition function in a more contained analytical expression thanks to the transfer matrix formalism (with exponentially small corrections in $N$)
\begin{equation}
    \log \mathcal Z_{\rm nn} = - \frac{\beta J N}{2} + N\log\!\left[\cosh(\beta\eps) + \sqrt{\sinh(\beta\eps)^2 +e^{2\beta J}}\right],
\end{equation}
from which we can compute analytically the expression for its derivatives and solve the geodesic equations efficiently with numerical tools. Let us remark here that since $\log\mathcal Z_{\rm nn}$ is linear in $N$ it is immediate that the same is true for its derivatives and therefore the dissipated work. Furthermore, it is quite clear from \eref{sm:kartoffel} that the differential equations will not depend on $N$, therefore the geodesic will also be independent of $N$.\\
\begin{figure}[H]
    \centering
    \includegraphics[width=.8\columnwidth]{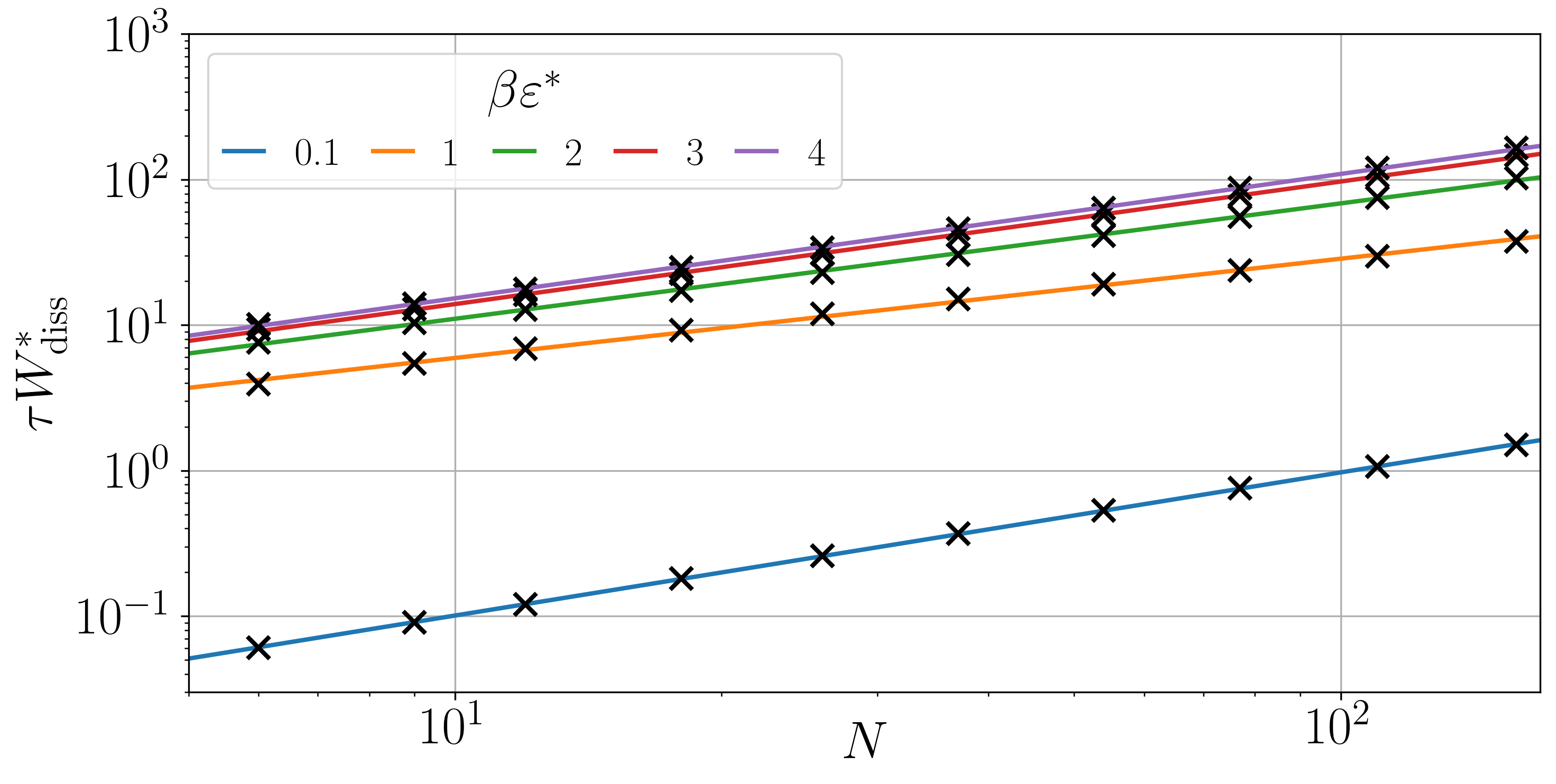}
    \vspace{-15pt}
    \caption{Scaling of the dissipation with respect to $N$ for different values of $\eps^*$.}
    \vspace{-15pt}
    \label{fig:scalings_sm}
\end{figure}
\begin{figure}[H]
    \centering
    \includegraphics[width=0.8\columnwidth]{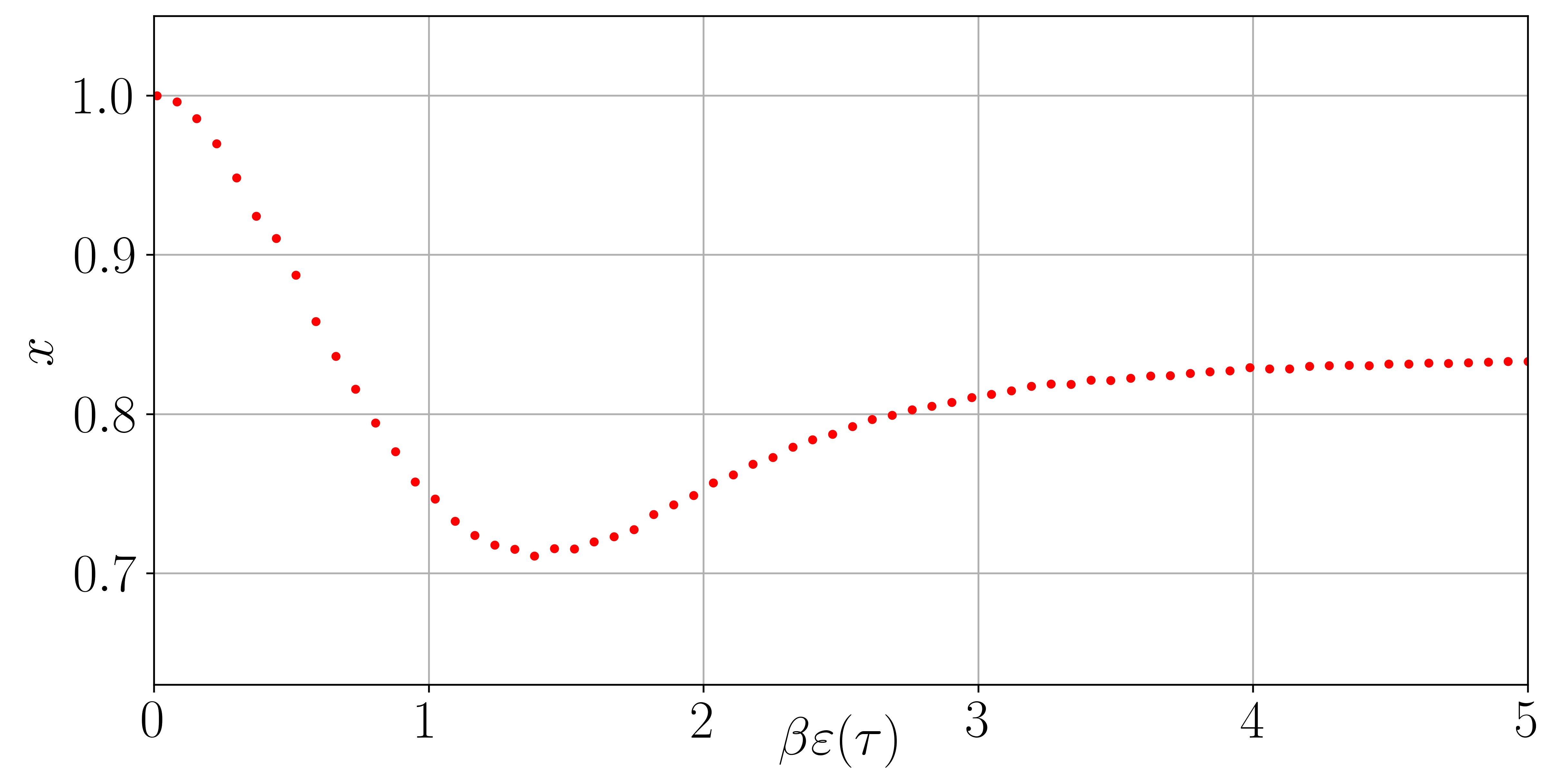}
    \vspace{-15pt}
    \caption{Dependence of the exponent $x$ of the dissipation $W_{\rm diss}^{\rm *, all} = \alpha N^x$ on the boundary condition $\beta\eps(\tau)$ (with $\eps(0) = 0$). The fit is achieved for numerical data up to $N=150$.}
    \label{fig:growth}
\end{figure}
For both models, we end up with a system of two second-order non-linear differential equations. By the structure of the geodesic equations, we can turn it into a system of four first-order non-linear differential equations of the type $\frac{d\vec x}{dt} = \vec f(\vec x)$ with $\vec x = (\eps, J, \dot\eps, \dot J)$. Here we want to enforce the boundary conditions $\eps(0) = J(0) = J(\tau)$ and $\eps(\tau) = \eps^*$ (where $\eps^* \gg k_B T$ for erasure), but famously boundary value problems are very difficult to solve numerically. In this particular case, we can exploit the fact that we have only two parameters and that geodesics never cross paths (they describe a flow in the parameter space). Therefore there is a bijection between the ratio of the initial velocities $\dot\eps(0)/\dot J(0)$ and the final value $\eps^*$, this allows us to turn the boundary value problem into solving multiple initial value problems for different ratios of the initial velocities until we find the initial conditions that match the desired boundary conditions.\\
Using these techniques we can find the geodesics for (approximate) erasure in both models, an example for each is showcased in \fref{fig:geo_all} and  \fref{fig:geo_1d}. As is explained in the main text, in the case of the all-to-all spin model the dissipation scales sub-linearly with respect to $N$. To quantify this effect we compute the dissipation for multiple values of $N$ (19 values for \fref{fig} and 10 values for \fref{fig:growth}, all between $N=5$ and $N=150$) and fit a power law. The relative errors of all the fits showcased in this study are $0.5\%$ or less. In \fref{fig:scalings_sm} we showcase more examples of this sub-linear scaling for different values of $\eps^*$ and how they are each well described by a power law. Then, in \fref{fig:growth} we collect all the fitted exponents to showcase the dependence on the boundary conditions. Where $x$ is plotted as a function of $\beta \eps(\tau)$, which illustrates that the collective effects developed here are genuinely process-dependent. 

\end{document}